\pdfoutput=1
\documentclass{report}

\usepackage{amsfonts,amsthm,amsmath,amssymb}
\usepackage{array}
\usepackage{epsfig}
\usepackage{color}
\usepackage{changebar}
\usepackage{url}
\usepackage{graphicx}
\usepackage{enumerate}
\usepackage{algpseudocode}
\usepackage{algorithmicx}
\usepackage[table]{xcolor}
\usepackage{pdflscape}
\usepackage{longtable}
\usepackage[margin=3cm]{geometry}

\newcommand{\example}[1]{\textbox{{\bf{Example:}}#1}}

\def\hicell{\cellcolor{green!25}}

\newcommand{\calI}{\mathcal{I}}
\newcommand{\calD}{\mathcal{D}}
\newcommand{\calH}{\mathcal{H}}
\newcommand{\calS}{\mathcal{S}}
\newcommand{\calU}{\mathcal{U}}

\newcommand{\R}{\mathbb{R}}

\newcommand{\N}{\mathbb{N}}

\newcommand{\veccc}[1]{\mathbf{#1}}
\newcommand{\matt}[1]{\mathbf{#1}}

\newcommand{\veczero}{\veccc{0}}
\newcommand{\vecone}{\veccc{1}}
\newcommand{\vecv}{\veccc{v}}
\newcommand{\vecd}{\veccc{d}}

\newcommand{\vecs}{\veccc{s}}
\newcommand{\vect}{\veccc{t}}

\newcommand{\vecx}{\veccc{x}}
\newcommand{\vecu}{\veccc{u}}

\newcommand{\matM}{\matt{M}}
\newcommand{\matI}{\matt{I}}
\newcommand{\matA}{\matt{A}}
\newcommand{\matC}{\matt{C}}

\def\matA{\matt{A}}

\def\vecr{\veccc{r}}

\def\vecz{\veccc{z}}

\def\vecs{\mathbf{s}}
\def\vecc{\mathbf{c}}

\newcommand{\str}		[1]	{\left\langle#1\right\rangle}

\newcommand{\set}		[1]	{\left\{#1\right\}}
\newcommand{\nin}			{\not\in}

\newcommand{\bigo}	[1]	{\mathcal{O}\left( #1 \right)}

\newcommand{\comment}	[1]	{}
\def\hinv{{\diamond -1}}
\newcommand{\vectwo}	[2]	{\left(
\begin{array}{c} 
#1 \\ #2 
\end{array}
\right)}

\newcommand{\textbox}	[1]	{\begin{center}\fbox{\parbox{15cm}{#1}}\end{center}}

\DeclareMathOperator*{\argmax}{\arg\!\max}

\graphicspath{ {./images/} }

\begin{document}

%%% This sets the page style and numbering for preliminary sections.
%\begin{preliminary}

\author{
David Kordalewski \\
Computer Science Theory Group \\
University of Toronto \\
Toronto, Canada \\
\texttt{kord@cs.toronto.edu}
}
\title{New Greedy Heuristics for Approximating Set Cover and Set Packing}
\date{April 2013}

%% This generates the title page from the information given above.
\maketitle

%% There should be NOTHING between the title page and abstract.
%% However, if your document is two-sided and you want the abstract
%% _not_ to appear on the back of the title page, then uncomment the
%% following line.
%\cleardoublepage

%% This generates the abstract page, with the line spacing adjusted
%% according to SGS guidelines.
\begin{abstract}
%% *** Put your Abstract here. ***
%% (At most 150 words for M.Sc. or 350 words for Ph.D.)

The Set Cover problem (SCP) and Set Packing problem (SPP) are standard NP-hard combinatorial optimization problems. Their decision problem versions are shown to be NP-Complete in Karp's 1972 paper. We specify a rough guide to constructing approximation heuristics that may have widespread applications and apply it to devise greedy approximation algorithms for SCP and SPP, where the selection heuristic is a variation of that in the standard greedy approximation algorithm. Our technique involves assigning to each input set a \emph{valuation} and then selecting, in each round, the set whose valuation is highest. We prove that the technique we use for determining a valuation of the input sets yields a unique value for all Set Cover instances. For both SCP and SPP we give experimental evidence that the valuations we specify are unique and can be computed to high precision quickly by an iterative algorithm. Others have experimented with testing the observed approximation ratio of various algorithms over a variety of randomly generated instances, and we have extensive experimental evidence to show the quality of the new algorithm relative to greedy heuristics in common use. Our algorithms are somewhat more computationally intensive than the standard heuristics, though they are still practical for large instances. We discuss some ways to speed up our algorithms that do not significantly distort their effectiveness in practice on random instances.

\end{abstract}

%% Anything placed between the abstract and table of contents will
%% appear on a separate page since the abstract ends with \newpage and
%% the table of contents starts with \clearpage.  Use \cleardoublepage
%% for anything that you want to appear on a right-hand page.

%% This generates a "dedication" section, if needed
%% (uncomment to have it appear in the document).
%\begin{dedication}
%% *** Put your Dedication here. ***
%\end{dedication}

%% The `dedication' and `acknowledgements' sections do not create new
%% pages so if you want the two sections to appear on separate pages,
%% you should put an explicit \newpage between them.

%% This generates an "acknowledgements" section, if needed
%% (uncomment to have it appear in the document).

%% This generates the Table of Contents (on a separate page).
\tableofcontents

%% This generates the List of Tables (on a separate page), if needed
%% (uncomment to have it appear in the document).
%\listoftables

%% This generates the List of Figures (on a separate page), if needed
%% (uncomment to have it appear in the document).
%\listoffigures

%% You can add commands here to generate any other material that belongs
%% in the head matter (for example, List of Plates, Index of Symbols, or
%% List of Appendices).

%% End of the preliminary sections: reset page style and numbering.
%\end{preliminary}

%%%%%%%%%%%%%%%%%%%%%%%%%%%%%%%%%%%%%%%%%%%%%%%%%%%%%%%%%%%%%%%%%%%%%%%%
%%  Put your Chapters here; the easiest way to do this is to keep     %%
%%  each chapter in a separate file and `\include' all the files.     %%
%%  Each chapter file should start with "\chapter{ChapterName}".      %%
%%  Note that using `\include' instead of `\input' will make each     %%
%%  chapter start on a new page, and allow you to format only parts   %%
%%  of your thesis at a time by using `\includeonly'.                 %%
%%%%%%%%%%%%%%%%%%%%%%%%%%%%%%%%%%%%%%%%%%%%%%%%%%%%%%%%%%%%%%%%%%%%%%%%

%% *** Include chapter files here. ***

\chapter{Introduction}
\label{chap:Introduction}

The Set Cover Problem (SCP) and Set Packing Problem (SPP) are standard NP-hard combinatorial optimization problems. Their decision problem versions are shown to be NP-Complete in \cite{karp}.
Because we cannot expect to solve all instances of these problems exactly in polynomial time, much effort has been expended on finding \emph{approximation} algorithms for these problems. Many algorithms have been proven to obtain approximate solutions for SCP and SPP that are within some factor of the optimal solution. At the same time there are results showing that, assuming some complexity conjectures, no polynomial-time algorithm can approximate these problems to any constant ratio. Further results have placed even stronger constraints on what sort of approximation ratios are achievable by polynomial-time algorithms.

Our primary purpose here is to describe a technique for greedily obtaining high-quality approximate solutions for Set Cover and Set Packing problems. Our technique involves assigning to each input set a \emph{valuation} and then selecting, in each round, the set whose valuation is highest. In order to specify the valuations, we define them recursively. For both SCP and SPP, we prove that a valuation satisfying our definition must exist, and in the case of Set Cover, that it is unique. We have not been able to show that our valuations result in a greedy algorithm that has some guaranteed approximation ratio, but do show experimentally that it performs somewhat better than the standard greedy algorithms for these problems on random instances. 

We believe that the mathematics underlying our recursively defined valuations and the overall idea that motivates our particular definitions are of significant interest independent of our algorithms and their performance.

%%%%%%%%%%%%%%%%%%%%%%%%%%%%%%%%%%%%%%%%%%%%%%%
\section{Notation and Terminology}
\label{sec:Notation and Terminology}
%%%%%%%%%%%%%%%%%%%%%%%%%%%%%%%%%%%%%%%%%%%%%%%
The notation $\vecd^\hinv$ is borrowed from \cite{fitzgerald1977fractional} to denote the \emph{Hadamard inverse} of a matrix or vector with all non-zero entries. If $\vecd$ is a vector, then the components of $\vecd^\hinv$ are $\left( \vecd^\hinv \right)_i = \frac 1 {\vecd_i}$.

We use the notation $A - B$ to denote the set difference of $A$ and $B$, $\set{x~|~ x\in A \wedge x \nin B}$.

We use $\matC$ to denote a diagonal matrix having $\vecc$'s entries on the diagonal.

For a matrix $\matM\in \R^{n\times n}$, we use $diag(\matM)$ to denote the length $n$ vector containing $\matM$'s diagonal entries.

A good proportion of the work on Set Cover uses $n$ for the number of input sets and $m$ for the size of the universe. In \cite{DBLP:journals/jacm/Feige98}, however, Feige uses $n$ for the universe size, as have some others since. By the rationale that $n$ should denote the quantity most deeply involved in attainable approximation ratios, we use $n$ for the universe size and $m$ for the number of input sets. 
%For other problems (Hitting Set, etc.), we use $n$ for the quantity corresponding to universe size when the problem is written as a Set Cover Instance.

The choice of using $\matA$ to denote a $n\times m$ matrix (rather than its transpose) was made simply so that the constraints in the integer program formulation of Set Cover would not require a transpose operation to be written.

We frequently use inequalities with vector quantities on either side, by which we intend the conjunction of the elementwise inequalities. For instance, $\vectwo{a}{b} \leq \vectwo{c}{d}$ should be understood as stating that $a\leq c$ and $b\leq d$.

We sometimes use the notation $\veczero$ or $\vecone$ to represent the column vector (of whatever size is appropriate in the context) with a 0 or 1 in every component, respectively.

%%%%%%%%%%%%%%%%%%%%%%%%%%%%%%%%%%%%%%%%%%%%%%%
\section{Organization}
%%%%%%%%%%%%%%%%%%%%%%%%%%%%%%%%%%%%%%%%%%%%%%%

In chapter \ref{chap:Problem Definitions and Discussion} we define the Set Cover and Set Packing problems, discuss some of the relationships that they have to other problems, and describe previous work on approximation algorithms for these problems.

Chapter \ref{chap:Preprocessing} outlines some preprocessing techniques that can be used to simplify SCP and SPP instances.

In chapter \ref{chap:New Greedy Heuristics for Set Cover and Set Packing}, we describe the overall idea that motivates our new heuristics and define the new heuristics themselves.

Chapter \ref{chap:Mathematical Results} contains our principal mathematical results, relating to vectors $\vecv\in\R^n_+$ for which $\matM\vecv=\vecv^\hinv$ for certain $n\times n$ matrices $\matM$.

Chapter \ref{chap:Numerical Calculation} describes a few algorithms that we have used effectively to compute our new valuations and includes some discussion of cases for which fixed points can be calculated exactly.

Chapter \ref{chap:Experimental Results} contains the results of experiments we have done, comparing the quality of approximate solutions to SCP and SPP instances obtained by a variety of simple greedy algorithms.

In a final brief chapter, we briefly summarize our results and suggest possible directions for future research.

%%%%%%%%%%%%%%%%%%%%%%%%%%%%%%%%%%%%%%%%%%%%%%%
\chapter{Problem Definitions and Discussion}
\label{chap:Problem Definitions and Discussion}
%%%%%%%%%%%%%%%%%%%%%%%%%%%%%%%%%%%%%%%%%%%%%%%

%%%%%%%%%%%%%%%%%%%%%%%%%%%%%%%%%%%%%%%%%%%%%%%
\section{Problem Formulations}
\label{sec:Problem Formulations}
%%%%%%%%%%%%%%%%%%%%%%%%%%%%%%%%%%%%%%%%%%%%%%%
\subsection{Set Cover}
%%%%%%%%%%%%%%%%%%%%%%%%%%%%%%%%%%%%%%%%%%%%%%%

An instance of the Set Cover Problem (SCP) has the following components:
\begin{enumerate}[(a)]
\item
	There is a set $\calI$. This contains the \emph{labels}, \emph{names} or \emph{indices} of the input sets and is only for notational convenience. This set can have arbitrary elements, but without loss of generality, we can take it to be $\set{1,\ldots,m}$.
\item
	$m=|\calI|$, the number of input sets.
\item
	For every $i\in \calI$, there is an input set $S_i$. The $S_i$'s can, again, be over arbitrary elements but without loss of generality, we can take it that $S_i \subseteq \set{1,\ldots,n}$.
\item
	$\calU = \bigcup_{i\in\calI} S_i$. $\calU$ is the universe or basis set of the instance.
\item
	$n = |\calU|$, the size of the universe.
\item
	Associated with every input $i \in \calI$ is a cost $c_i\in\R$.
\end{enumerate}
%Given the above basic components of a Set Cover instance, we can define some convenient quantities derived from them. 
%We will use these quantities as shorthand for their definitions. In particular, when describing algorithms 

%\begin{enumerate}[(a)]
%\item
	%There is a set $\calU$. We define $n = |\calU|$. This is the \emph{universe} or \emph{basis} set that is used. Typically, we will use $\calU = \set{1,\ldots,n}$ for convenience.
%\item
	%$\calS = \set{s_1,\ldots,s_m}$ is a collection of $m$ sets over $\calU$, so that for all $1\leq i \leq m$ we have $s_i \subseteq \calU$. 
%\item
	%Associated with every input set $s_i$ is a cost $c_i\in\R$.
%\end{enumerate}

We say that $\calH \subseteq \calI$ covers $\calU$ or that $\calH$ is a cover for $\calU$ when $\bigcup_{i\in \calH} S_i = \calU$. That is, every basis element is included in at least one of $\calH$'s sets. 
Define the cost of any $\calH \subseteq \calI$ to be $\sum_{i \in \calH} c_i$, the sum of the costs of the sets included in $\calH$.

The Set Cover Problem is to find, given $\calI$, $S_i$ and $c_i$ for $i \in \calI$, a set cover with minimal cost. The Unweighted Set Cover Problem describes instances for which $c_i=1$ for all input sets $i$. The decision problem variant of Set Cover is to determine, given $\calI$, $S_i$ and $c_i$ for $i \in \calI$ and a cost threshold $k$, whether there is a cover $\calH \subseteq \calI$ with cost not exceeding $k$.

%%%%%%%%%%%%%%%%%%%%%%%%%%%%%%%%%%%%%%%%%%%%%%%
\subsection{Hitting Set}
%%%%%%%%%%%%%%%%%%%%%%%%%%%%%%%%%%%%%%%%%%%%%%%
Hitting Set is another common NP-hard combinatorial optimization problem.
An instance of Hitting Set has the following components.
\begin{enumerate}[(a)]
\item
	There is a set $\calU$. As in Set Cover, we call these the \emph{labels} or \emph{names} of the input sets. 
\item
	Define $n = |\calU|$.
\item
	For every $i\in \calU$, there is an input set $S_i$.
\item 
	Define $\calI = \bigcup_{i\in\calU} S_i$. 
\item
	Define $m=|\calI|$.
\item
	Associated with every element $i \in \calI$ is a cost $c_i\in\R$.
\end{enumerate}

We will say that $\calH \subseteq \calI$ hits an input set $S_i$ if $|\calH \cap S_i| \geq 1$, and that $\calH \subseteq \calI$ is a hitting set for $\calU$ if for every $i \in \calU$, $\calH$ hits $S_i$.
Define the cost of any $\calH \subseteq \calI$ to be $\sum_{i \in \calH} c_i$, the sum of the costs of the basis elements included in $\calH$.

The Hitting Set problem is, then, to find a minimum cost hitting set $\calH \subseteq \calI$ for $\calU$.

\subsection{(Hitting \{Set) Cover\} }
%%%%%%%%%%%%%%%%%%%%%%%%%%%%%%%%%%%%%%%%%%%%%%%
The construction given below is commonly used to show the equivalence of Hitting Set and Set Cover, but here we use it as a problem definition.
We regard this problem as unifying Hitting Set and Set Cover into a uniform terminology that easily translates back to either problem. For lack of a standard name, we will call this problem (Hitting \{Set) Cover\} or HSC.

An HSC instance consists of the following objects. 
\begin{enumerate}[(a)]
\item
	$L = \set{l_1,\ldots,l_m}$ and $R = \set{r_1,\ldots,r_n}$ are disjoint sets. 
\item
	Let $c:L \to \R$ give the costs of $L$'s elements. 
\item
	$G = \str{ L \cup R, E }$ is an undirected bipartite graph where $E\subseteq L \times R$, so every edge connects one element of $L$ with one from $R$. 
\end{enumerate}

We then define a hitting set cover for the problem $(G,c)$ to be any $\calH \subseteq L$, the union of whose neighbours is $R$. That is, for every right element $r \in R$ there is some $l \in \calH$ so that the edge $(l,r) \in E$ exists. For any subset $\calH \subseteq L$, we define its cost $c(\calH)$ as the sum of its elements' costs $c(\calH)=\sum_{l\in \calH} c(l)$. A minimum hitting set cover, then, is any hitting set cover whose cost is least possible among all hitting set covers.

To see why this problem unifies hitting set and set cover, consider the adjacency matrix of $G$. Since $G$ is bipartite, there are no edges between pairs of elements both in either of $L$ or $R$, so we can write its adjacency matrix as
\[
\matM_G=
\left( \begin{array}{cc}
0 & \matA  \\
\matA^T & 0 \end{array} \right)
\]
where $\matA$ is an $n\times m$ matrix with 
\[
\matA_{i,j}=
   \left\{
     \begin{array}{lr}
       1 & \text{ if } (l_j, r_i) \in E \\
       0 & o.w.
     \end{array}
   \right.
\]

We will call $\matA$ the {fundamental matrix} of a HSC problem. We would also like to define $\vecc$ as the vector of costs of $L$'s elements, so that for $i \in \set{1,\ldots,m}$, we have $\vecc_i=c(l_i)$.

Now, $\vecc$ and $\matA$ contain a complete description of our problem. When the columns of $\matA$ are set to the adjacency vectors of the input sets from a hitting set instance, we have a problem in this setting equivalent to the hitting set problem. Likewise, if we set $\matA$'s rows to be the input sets from an instance of Set Cover, we have an equivalent problem. Because we can always transform instances of hitting set and set cover into problems of this sort, we can treat both problems uniformly by considering this problem instead.
%Intuitively, we can see this problem description containing simultaneously the (in this problem's terminology) $L$-centric description of Set Cover and the $R$-centric description of Hitting Set. This and the Integer Program formulation below are the standard ways we discuss the Set Cover Problem when we choose to depart from the more commonly understood notions of the basic Set Cover Problem definition.

As an example, consider the SCP instance $\calI = \set{1,2,3,4,5}$, with $S_1=\set{1,2,3}$, $S_2=\set{2,4}$, $S_3=\set{1,3}$, $S_4=\set{4}$, and $S_5=\set{3,4}$. The costs are given by $c_1=3, c_2=1,c_3=2,c_4=1,c_5=2$. The figure below shows the HSC instance that this generates.

\begin{figure}[!ht]
\centering
\includegraphics[width=0.4\textwidth]{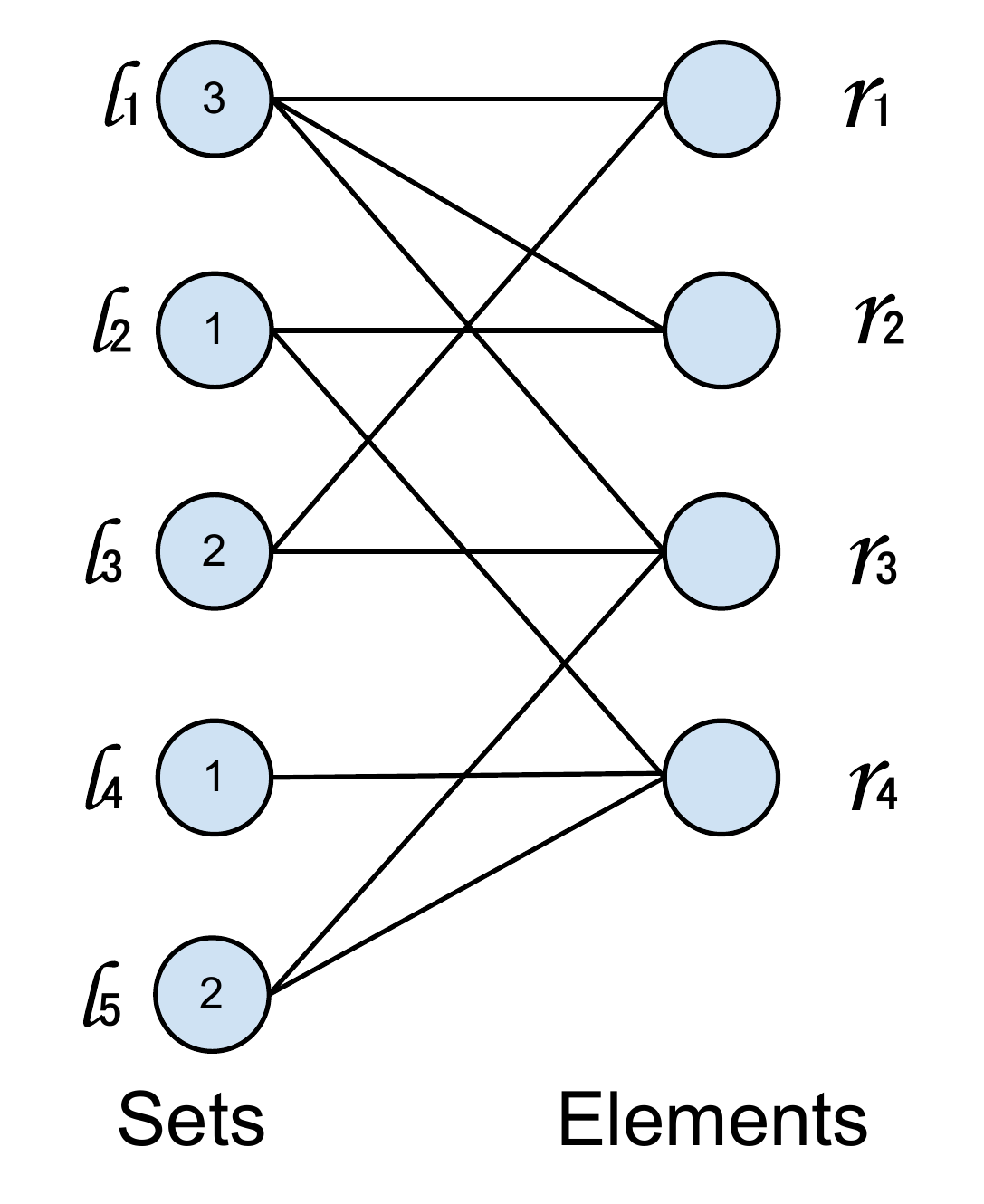}
\caption{Graph structure of an HSC instance.}
\label{fig:hitting-set-cover-example-instance}
\end{figure}

If we were to take the Hitting Set instance $\calU=\set{1,2,3,4}$, with $S_1=\set{1,3}$, $S_2=\set{1,2}$, $S_3=\set{1,3,5}$, $S_4=\set{2,4,5}$ and costs $c_1=3, c_2=1,c_3=2,c_4=1,c_5=2$, we obtain the same equivalent HSC instance.

This instance has fundamental matrix and cost vector given by the following:
\begin{align*}
\matA = 
\left(
\begin{array}{ccccc}
 1 & 0 & 1 & 0 & 0 \\
 1 & 1 & 0 & 0 & 0 \\
 1 & 0 & 1 & 0 & 1 \\
 0 & 1 & 0 & 1 & 1 \\
\end{array}
\right)
&~~~
\vecc = 
\left(
\begin{array}{c}
 3 \\
 1 \\
 2 \\
 1 \\
 2 \\
\end{array}
\right)
\end{align*}

%%%%%%%%%%%%%%%%%%%%%%%%%%%%%%%%%%%%%%%%%%%%%%%
\subsection{Integer Programming Formulation}
%%%%%%%%%%%%%%%%%%%%%%%%%%%%%%%%%%%%%%%%%%%%%%%
Set Cover can be written very simply as an equivalent integer program, in which:
\begin{enumerate}[(a)]
\item
	$\matA$ is some particular $n\times m$ $0/1$ matrix.
\item
	$\vecc$ is a vector of costs: $\vecc \in \R^m$.
\item
	$\vecx$ is a vector of binary variables.
\end{enumerate}
Then the IP problem is
\begin{align*}
&\text{Minimize } \vecc^T \vecx
\\
&\text{subject to } \matA\vecx \geq \vecone
\text{ and }
\vecx_i \in \set{0,1} \text{ for all } i=1,\ldots,m
\end{align*}
This problem is equivalent to the SCP instance whose subsets are given by understanding the columns of $\matA$ as adjacency vectors that indicate which of the $n$ basis elements are included in each input set. This is equivalent to the hitting set instance whose sets to hit are given by $\matA$'s rows.

%%%%%%%%%%%%%%%%%%%%%%%%%%%%%%%%%%%%%%%%%%%%%%%
\subsection{Set Packing and Formulations}
%%%%%%%%%%%%%%%%%%%%%%%%%%%%%%%%%%%%%%%%%%%%%%%

Set Packing is also a well known NP-hard combinatorial optimization algorithm. 
As with SCP, a Set Packing instance consists of the following objects:
%\begin{enumerate}[(a)]
%\item
	%There is a set $\calU$. We define $n = |\calU|$. This is the \emph{universe} or \emph{basis} set that is used. Typically, we use $\calU = \set{1,\ldots,n}$ for convenience.
%\item
	%$\calS = \set{s_1,\ldots,s_m}$ is a collection of $m$ sets over $\calU$, so that for all $1\leq i \leq m$ we have $s_i \subseteq \calU$. 
%\item
	%Associated with every input set $s_i$ is a weight $c_i\in \R$.
%	\footnote{The word ``weight'' is used instead of ``cost'' here due to the convention that quantities to be maximized are weights, while quantities to be minimized are costs.}
%\end{enumerate}
\begin{enumerate}[(a)]
\item
	There is a set $\calI$. This contains the \emph{labels} or \emph{names} of the input sets.
\item
	Define $m = |\calI|$.
\item
	For every $i\in \calI$, there is an input set $S_i$. 
\item
	Define $\calU = \bigcup_{i\in \calI} S_i$. $\calU$ is the universe or basis set of the problem.
\item
	Define $n = |\calU|$.
\item
	Associated with every input set $i \in \calI$ is a cost $c_i\in\R$
\end{enumerate}
Call $\calH \subseteq \calI$ a packing for $\calU$ if every basis element is included in at most one of $\calH$'s sets. 
Equivalently, $\calH \subseteq \calI$ is a packing for $\calU$ if the input sets selected are pairwise disjoint. That is, $\forall i, j \in \calH, i = j \text{ or } S_i \cap S_j = \emptyset$. 
Also define the weight of a packing $\calH \subseteq \calI$ to be the sum of the weights of the included sets, $c(\calH) = \sum_{i \in \calH} c_i$.
The Set Packing problem is to find the maximum weight packing $\calH \subset \calS$ of $\calU$.

It can be seen that the description of a Set Cover instance has precisely the same description as a Set Packing instance. Given $\calI, S_i, c_i$ for $i \in \calI, $, we can just as well ask what the maximum weight packing or the minimum cost cover is.

Much the same as with the relationship between Set Cover and Hitting Set, Set Packing has an analogous equivalent problem phrased in terms of elements with costs and constraints on subsets of these elements. This problem has not been given any particular attention that we are aware of, but we name and define it here for the sake of strengthening the analogy between Set Cover and Set Packing. We call this the \emph{Jabbing Set Problem}.

A Jabbing Set instance consists of the following objects:
\begin{enumerate}[(a)]
\item
	There is a set $\calU$. This contains the \emph{labels} or \emph{names} of the input sets. 
\item
	Define $n = |\calU|$.
\item
	For every $i \in \calU$, there is an input set $S_i$. 
\item
	Define $\calI = \bigcup_{i\in \calU} S_i$. $\calI$ is the universe or basis set of the problem.
\item
	Define $m = |\calI|$.
\item
	Associated with every input $i \in \calI$ is a cost $c_i\in\R$
\end{enumerate}
Call $\calH \subseteq \calI$ a jabbing set for $\calU$ if every subset $S_i$ for $i \in \calU$ is hit at most once, $\forall i\in \calI, |S_i \cap \calH| \leq 1$. 
The Jabbing Set problem is then to find the maximum weight jabbing set $\calH \subseteq \calI$ for $\calU$.

Much as with Set Cover, we will often favor the Integer Program formulation of Set Packing.
Set Packing can be written very simply as an integer program as follows, differing from the formulation for Set Cover only by the direction of an inequality and swapping the \emph{minimize} for \emph{maximize}:
\begin{enumerate}[(a)]
\item
	$\matA$ is some particular $n\times m$ $0/1$ matrix.
\item
	$\vecc$ is a vector of costs: $\vecc \in \R^m$.
\item
	$\vecx$ is our vector of binary variables.
\end{enumerate}
Then the IP problem is
\begin{align*}
&\text{Maximize } \vecc^T \vecx
\\
&\text{subject to } \matA\vecx \leq \vecone
\text{ and }
\vecx_i \in \set{0,1} \text{ for all } i=1,\ldots,m
\end{align*}

%%%%%%%%%%%%%%%%%%%%%%%%%%%%%%%%%%%%%%%%%%%%%%%
\subsection{Set Packing's Relationship to Maximum Independent Set}
\label{sec:Set Packing's Relationship to Maximum Independent Set}
%%%%%%%%%%%%%%%%%%%%%%%%%%%%%%%%%%%%%%%%%%%%%%%
The Weighted Maximum Independent Set (WMIS) problem (also called Vertex Packing) can be described as follows: For a graph $G=\str{V,E}$ with vertex weights given by $c_i$ for $i \in V$, what is the largest weight subset $\calH$ of $V$ such that no pair of elements from $\calH$ are neighbours in $G$. This problem is also NP-hard.

The reductions between Set Packing and WMIS are particularly clean. Given a WMIS problem defined by $G=\str{V,E}$ and $c$, an equivalent Set Packing instance is $\calI=V$, $S_i=\set{e\in E~|~ i \in e}$. Any max weight independent set $\calH \subseteq V$ for $G, c$ is also a max weight packing for $V, c$ and the $S_i$'s. To see this construction more clearly, consider the elements' neighbourhoods for the SPP instance. For every $e\in\calU$ we can define $N_e = \set{i\in\calI ~|~ e\in S_i}$. Then, for every $e\in E$, we make an element with a neighbourhood of size two.

Given instead some Set Packing instance $\calI, c$ and $S_i$ for $i\in\calI$, we can make an equivalent WMIS instance by defining $G=\str{\calI,E}$ where $E$ has an edge $(i,j)$ for every $i\neq j$ in $\calI$ iff $S_i\cap S_j \neq \emptyset$. Then the optimal solutions to these problems are the same. 
By considering the element neighbourhoods, we can construct the WMIS problem even more simply. For each $e\in\calU$ we add a clique to $G$, yielding an edge between every $i,j \in N_e$. 

Viewed this way, we can see SPP as an alternative way to specify WMIS instances where we can write a graph as a union of arbitrary size cliques, rather than just 2-cliques, which are simply edges. This will be relevant later in section \ref{sec:The New Greedy Set Packing Heuristic}, where we specify our new Set Packing heuristic.

To be more formal, consider a problem with input $V$ some set, $S_1,\ldots,S_n$ all subsets of $V$ specifying cliques among $V$'s elements and $c:V\to \R$ weights of the elements of $V$. The problem is to find the WMIS of $G=\str{V,E}$ with weights given by $c$ and $E = \bigcup_{i,j\in\set{1,\ldots,n}, i\neq j \text{ and }S_i\cap S_j\neq \emptyset} \set{(i,j)}$. If we restricted this to instances where $|S_i|=2$ for all $i\in\set{1,\ldots,n}$, the problem is WMIS. Without that restriction, this problem is equivalent to Set Packing in precisely the same way that Set Cover and Hitting Set are equivalent.

%%%%%%%%%%%%%%%%%%%%%%%%%%%%%%%%%%%%%%%%%%%%%%%
%\subsection{$k$-Set Packing}
%%%%%%%%%%%%%%%%%%%%%%%%%%%%%%%%%%%%%%%%%%%%%%%

%$k$-Set Packing is the Set Packing problem where all input sets are restricted to have no more than $k$ elements. Many results about Set Packing are 

%\Undone

%%%%%%%%%%%%%%%%%%%%%%%%%%%%%%%%%%%%%%%%%%%%%%%
\section{Greedy Algorithms for Set Cover/Packing Approximation}
%%%%%%%%%%%%%%%%%%%%%%%%%%%%%%%%%%%%%%%%%%%%%%%

%%%%%%%%%%%%%%%%%%%%%%%%%%%%%%%%%%%%%%%%%%%%%%%
\subsection{Set Cover/Packing Approximation}
\label{sec:Set Cover/Packing Approximation}
%%%%%%%%%%%%%%%%%%%%%%%%%%%%%%%%%%%%%%%%%%%%%%%

Since the underlying problems are NP-hard, and are not expected to have polynomial time algorithms for exact solution, it has become an area of interest to find algorithms that run in polynomial time and perform well by some measure. Researchers have been particularly interested in algorithms that offer a guarantee on the ratio between the provided solution and the optimal solution. The standard greedy algorithm for SCP does just that, and provides a worst-case approximation ratio very close to what has been proven (under plausible assumptions) the best possible worst-case approximation ratio for any polynomial time algorithm.

%%%%%%%%%%%%%%%%%%%%%%%%%%%%%%%%%%%%%%%%%%%%%%%
\subsection{Set Cover Approximation}
\label{sec:Set Cover Approximation}
%%%%%%%%%%%%%%%%%%%%%%%%%%%%%%%%%%%%%%%%%%%%%%%

%\workingnote{I would like to have algorithms named and numbered nicely, but my document fails to compile when I include the appropriate package. I'll have to hack through this sooner or later.}

The best studied technique for obtaining good approximate solutions to Set Cover problems was first discussed in \cite{chvatal1979greedy}. Let the input be a $\calI$, the sets $S_i$ for all $i\in\calI$ and costs $c_i\in\R_+$ for every $i \in \calI$. We can describe the \emph{Standard Greedy Algorithm for SCP} in the following pseudocode. 

%\begin{algorithm}
%\caption{Standard Greedy SCP Algorithm}
\begin{algorithmic}
%\If{$\bigcup_{s\in \calS}s \neq \calU$}
%	\State Problem is infeasible
%\EndIf
\State $\calH \gets \emptyset$
\While{$\bigcup_{i\in \calI} S_i \neq \emptyset$} 
	\State $b \gets \argmax_{i \in \calI} \frac{|S_i|}{c_i}$
	\State $\calH \gets \calH \cup \set{b}$
	\ForAll{$i \in \calI$}
		\State $S_i \gets S_i - S_b$
	\EndFor
\EndWhile
\State\Return $\calH$
\end{algorithmic}
%\end{algorithm}

Here, $\calH$ contains the indices of the sets that have been selected so far to form a cover of $\calU$.  In every run through the while loop, the algorithm selects the set maximizing the ratio between the number of uncovered elements and the cost of the set. It then modifies the input sets to remove the newly covered elements from them, effectively removing some elements of $\calU$ since they have already been covered and it is irrelevant whether they are covered again or not.

%%%%%%%%%%%%%%%%%%%%%%%%%%%%%%%%%%%%%%%%%%%%%%%
\subsection{Set Packing Approximation}
\label{sec:Set Packing Approximation}
%%%%%%%%%%%%%%%%%%%%%%%%%%%%%%%%%%%%%%%%%%%%%%%

For Set Packing, we describe an analogous algorithm that greedily selects additional sets until no more can be added. For convenience, we describe the weights somewhat differently than in the above. Let the input be a set of input set names $\calI$, each of the sets $S_i$ for $i \in \calI$, and weights given by $c_i\in\R_+$ for $i \in \calI$

\begin{algorithmic}
%\If{$\bigcup_{s\in \calS}s \neq \calU$}
%	\State Problem is infeasible
%\EndIf
\State $\calH \gets \emptyset$
\While{$\calI \neq \emptyset$}
	\State $b \gets \argmax_{i\in \calI} \frac{c_i}{\sqrt{|S_i|}}$ \Comment{Here we can use $\frac{c_i^2}{|S_i|}}$ instead for the same results.
	\State $\calH \gets \calH \cup \set{b}$
	\ForAll{$i \in \calI$}
		\If{$S_i \cap S_b \neq \emptyset$}
			\State $\calI \gets \calI - i$
		\EndIf
	\EndFor
\EndWhile
\State\Return $\calH$
\end{algorithmic}

In this algorithm, we select the remaining feasible set for which the cost per square root of the number of elements is largest. We then remove all sets that contain any element in common with the selected set, effectively removing some sets from $\calS$ if our latest choice means that their selection would be infeasible. The particular choice of the set maximizing $c_i / \sqrt{|S_i|}$ is similar to the algorithm for Set Cover described above, though the square root of the set size may seem odd. In fact, this algorithm has the optimal worst-case approximation ratio for any greedy set packing algorithm in which the valuation of every set $i$ is determined by $|S_i|$ and $\vecc_i$ and so does  not involve the detailed structure of the instance \cite{gonen2000optimal}.
In section~\ref{sec:Other Approximation Techniques} we will discuss this and other interesting heuristics.

%%%%%%%%%%%%%%%%%%%%%%%%%%%%%%%%%%%%%%%%%%%%%%%
\subsection{General Greedy Scheme}
\label{sec:General Greedy Scheme}
%%%%%%%%%%%%%%%%%%%%%%%%%%%%%%%%%%%%%%%%%%%%%%%

The most important line in each of the above algorithms is where the next set to use is chosen. We would like to describe a general greedy algorithm for approximating Cover and Packing problems, abstracting away the particular rationale for selection. 

The previous two algorithms provide some \emph{valuation} of the sets in the instance, assigning a real number to each of them indicating their relative desirability. Both also \emph{reduce} the underlying instance to reflect the fact that some set has been selected. How this reduction is performed depends on whether we are dealing with a Cover or Packing problem.
In addition to these, we can \emph{transform} or \emph{preprocess} the instance to a simpler one when the instance has some simple properties. One example suggesting the type of preprocessing we mean is the situation where (for SCP) some input set $S_i$ is a subset of $S_j$ and $c_i>c_j$, so that we immediately know that $S_i$ will not be in any optimal cover since $S_j$ is both larger and cheaper and we can remove $S_i$ from the problem entirely. We will discuss this in greater detail in chapter \ref{chap:Preprocessing}.

The overall scheme can be described in pseudocode as follows, where $P$ denotes an instance:

\begin{algorithmic}
%\If{$\bigcup_{s\in \calS}s \neq \calU$}
%	\State Problem is infeasible
%\EndIf
\State $preprocess(P)$
\While{a choice can or must be made}
	\State $\vecv \gets valuation(P)$
	\State $b \gets \argmax_{i \in \calI} \vecv_i$
	\State $\calH \gets \calH \cup {b}$
	\State $reduce(P, b)$
	\State $preprocess(P)$
\EndWhile
\State\Return $\calH$
\end{algorithmic}

The preprocessing step is not usually included in a description of the standard greedy algorithm for Set Cover, but it can make a difference in the performance of the algorithm. Usually, however, it is not seen as important whether preprocessing is done on SCP instances where the standard greedy algorithm is concerned, as it does not impact the worst-case approximation ratio. For the valuation technique we describe later, certain types of preprocessing are relatively cheap and have an impact on the valuations, whereas for the standard greedy algorithm they do not. This is discussed further in chapter \ref{chap:Preprocessing}.

%%%%%%%%%%%%%%%%%%%%%%%%%%%%%%%%%%%%%%%%%%%%%%%
\section{The Standard Greedy Cover Heuristic}
\label{sec:The Standard Greedy Heuristic}
%%%%%%%%%%%%%%%%%%%%%%%%%%%%%%%%%%%%%%%%%%%%%%%

The standard greedy heuristic for Set Cover is to choose the set for which the number of uncovered elements per unit cost is largest. It was first described in \cite{johnson1974approximation} for unweighted instances. For a Set Cover instance given by the matrix $\matA$ and element costs $\vecc$, as in the IP formulation of the problem, the valuation $\vecv_i$ of set $S_i$ corresponding to column $i$ of $\matA$ will be 
\[
\vecv_i = \sum_{j=1}^n \matA_{j,i} / \vecc_i = \frac{1}{\vecc_i} \sum_{j=1}^n \matA^T_{i,j}
\]
If we denote the $m\times m$ diagonal matrix with diagonal entries given by $\vecc_i^{-1}$ for $1\leq i \leq m$ by $\matC^{-1}$, we can write the entire valuation vector simply as
\[
\vecv = \matC^{-1} \matA^T \vecone
\]

In \cite{chvatal1979greedy}, Chv\`atal demonstrates that the approximate solutions for Set Cover problems obtained using this valuation must have have a cost of no more than $H(n)$ times the optimal solution, where $H(n)=\sum_{i=1}^n \frac 1i$ is the sum of harmonic series up to the $n$th term. In fact, Chv\`atal showed that this approximation ratio cannot exceed $H(k)$, where $k = \max_{i\in \calI} |S_i|$, the size of the largest input set, which obviously cannot exceed $n$.

\begin{figure}[!ht]
\centering
\includegraphics[width=0.5\textwidth]{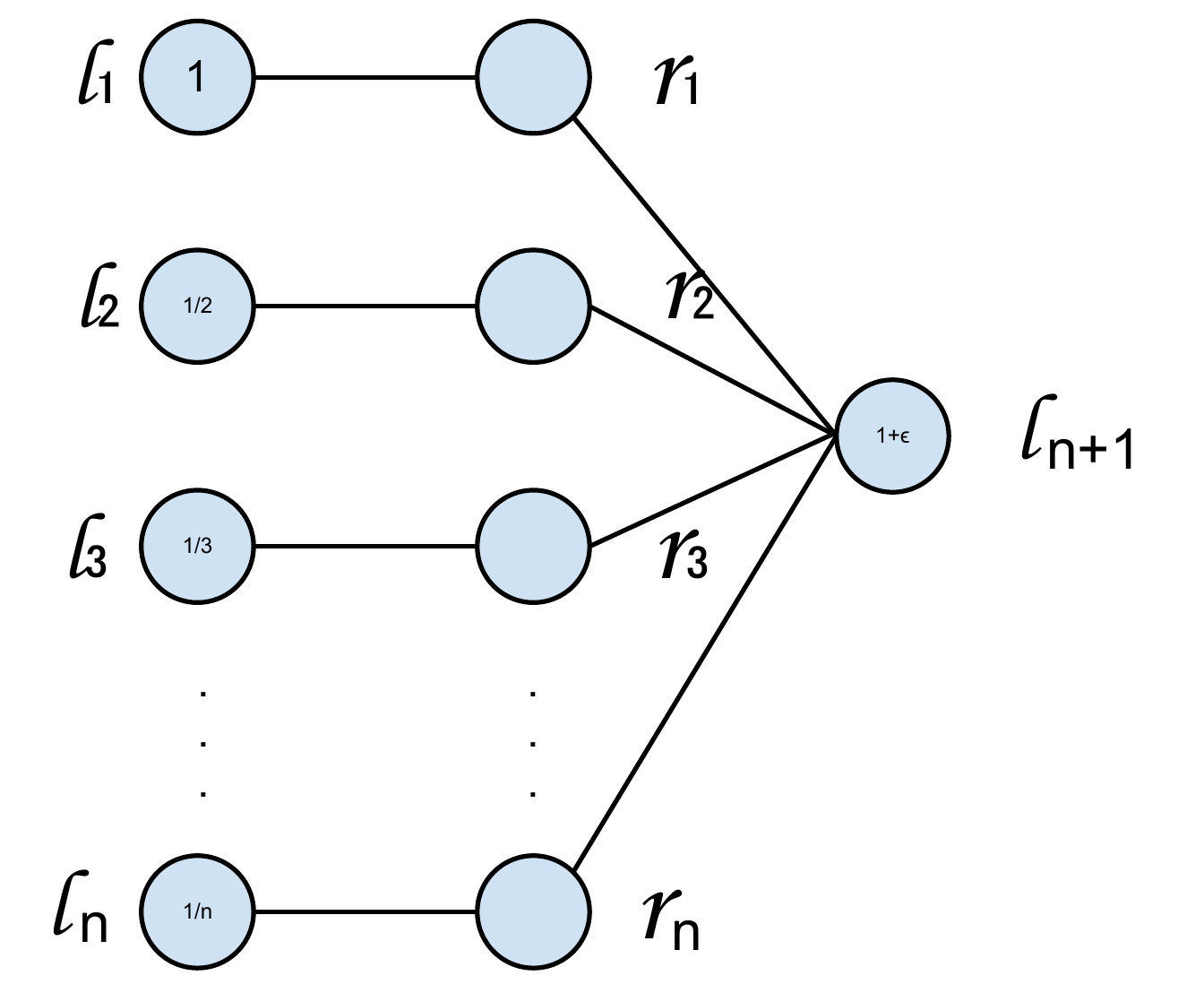}
\caption{Graph structure of hard Instances for standard greedy set cover. The costs of $l_i$ for $1\leq i\leq n$ are $c_i=1/i$ and the cost of $l_{n+1}$ is $c_{n+1}=1+\epsilon$}
\label{fig:standard_hard_instance}
\end{figure}

It is also known that for all $n$, there are instances for which an approximation arbitrarily close to $H(n)$ is attained. We can give a class of such instances quite simply. Let $\calI = \set{1,\ldots,n+1}$ and for every $1\leq i\leq n$ let $S_i = \set{r_i}$ and $S_{n+1} = \set{r_1,\ldots,r_n}$. Figure \ref{fig:standard_hard_instance} shows the HSC problem graph structure to which this corresponds. The costs of the input sets for $1 \leq i \leq n$ are $c_i=1/i$, and the cost of the last set $c_{n+1}=1+\epsilon$ for any $\epsilon > 0$.

Now, for this instance, the valuation that the standard greedy heuristic provides is 
\[
\left(
1,
2,
3,
\ldots,
i, 
\ldots,
n, 
\frac {n}{1+\epsilon}
\right)^T
\]
Since the largest of these values is $n$, we will take the set corresponding to it, $S_n$, incurring cost $1/n$ and remove $S_n=\set{r_n}$ from every set. In the next iteration, we are given the valuation $\left(1,2,3,\ldots, i, \ldots ,n-1, \frac {n-1}{1+\epsilon} \right)^T$ and select $S_{n-1}$. This will continue until the final set cover produced is $\calH=\set{1,\ldots,n}$ with cost $H(n)$. The optimal set cover, however, is $OPT=\set{n+1}$ so the attained approximation ratio is $\frac{c(\calH)}{c(OPT)}=\frac{H(n)}  {1+\epsilon}$ and since epsilon can be arbitrarily small, these instances have approximation ratios arbitrarily close to $H(n)$.

The performance of this heuristic for solving Unweighted Set Cover problems, where all costs $c_i$ are fixed at 1, is also well understood. In \cite{DBLP:dblp_conf/stoc/Slavik96}, Slav\'ik shows that the standard greedy heuristic provides solutions for unweighted instances whose cost is within a factor of $\ln(n) - \ln(\ln( n)) + 0.78$ of the cost of the optimal solution. He also shows that this ratio is tight. For every $n\geq 2$, there is an instance with $|\calU|=n$ and the standard greedy heuristic yields a solution with cost more than $\ln (n) - \ln (\ln (n)) -0.31$ times the optimal solution's cost. Note that this is better than the worst-case performance over weighted Set Cover instances, since $H(n) \approx \ln (n) + 0.58$.

%%%%%%%%%%%%%%%%%%%%%%%%%%%%%%%%%%%%%%%%%%%%%%%
\section{Other Approximation Techniques}
\label{sec:Other Approximation Techniques}
%%%%%%%%%%%%%%%%%%%%%%%%%%%%%%%%%%%%%%%%%%%%%%%

A variety of non-greedy poly-time algorithms have been proposed and used for approximating SCP. LP rounding, dual LP rounding, and primal/dual have been described and proven to have good worst-case approximation ratios. Williamson in \cite{williamson2002primal}, in particular, is a good resource for more information on the primal/dual technique.

In \cite{halldorsson1995approximating} Halld{\'o}rsson describes an algorithm using local search that has worst-case approximation ratio better than the standard greedy algorithm by a constant. Beasley and Chu describe a genetic algorithm-based approach in \cite{chu1998genetic}.
In \cite{grossman1997computational}, Grossman and Wool describe some variations of the standard greedy algorithm in addition to a neural network-based algorithm.
Caprara et al.\ give a Langrangian-based algorithm in \cite{caprara1999heuristic} that won the 1994 Set Cover approximation competition \emph{FASTER}.

%%%%%%%%%%%%%%%%%%%%%%%%%%%%%%%%%%%%%%%%%%%%%%%
%\section{Known Results For Cover/Packing Approximation}
%\label{sec:Known Results For Cover/Packing Approximation}
%%%%%%%%%%%%%%%%%%%%%%%%%%%%%%%%%%%%%%%%%%%%%%%

%In addition to the 

\chapter{Preprocessing}
\label{chap:Preprocessing}

Here we wish to describe some particular simplifications of Set Cover/Packing instances that can be done cheaply and result in problems that are equivalent to and no larger than the input problems. Any greedy algorithm utilizing the scheme presented in section \ref{sec:General Greedy Scheme} with some guaranteed approximation ratio that is non-decreasing in $n$ and $m$ (and perhaps $k=\max_{i\in\calI} |S_i|$) cannot have a worse worst-case approximation ratio when these preprocessing steps are used.

We discuss some ways to take an instance $I$ (in some convenient representation) and produce a new instance $I'$ such that any optimal solution to $I'$, possibly taken with some specified sets that must be included, is an optimal solution to $I$ and $n' \leq n$, $m'\leq m$ and $d' \leq d$. 
%We will also be concerned somewhat with a notion of equivalence. If any pair of instances $I$ and $J$ have the property that there is some instance $K$ such that some series of preprocessing steps applied to $I$ result in $K$ and that some series of preprocessing steps applied to $J$ also result in $K$.

For transformations satisfying the above, it should generally be beneficial to perform the transformations and solve the transformed problems as opposed to solving the original problem with the same algorithm. However, it is certainly not the case that all algorithms for approximating SCP can solve the transformed problem with approximation ratio no worse than the original problem for all instances. Consider a transformation as unproblematic as reordering the basis elements, which could just as likely help as hinder an algorithm with some deterministic tie-breaking procedure. Nor can it be said that the approximation ratio must always be improved for all approximation algorithms. All we wish to claim for transformations of this sort is that they cannot worsen the worst-case approximation ratio of the greedy valuation technique used and they frequently improve the approximation ratio when applied.

%%%%%%%%%%%%%%%%%%%%%%%%%%%%%%%%%%%%%%%%%%%%%%%
\section{Basic Preprocessing}
\label{sec:Basic Preprocessing}
%%%%%%%%%%%%%%%%%%%%%%%%%%%%%%%%%%%%%%%%%%%%%%%
First, the following are a few ways in which the resulting instances are very clearly equivalent. 
\begin{enumerate}[(a)]
\item
Renaming the elements of the universe results in a problem that is semantically equivalent, though its representation can be somewhat different. 

\item
Renaming the input set labels.

\item Scaling all of the costs by some constant $\lambda > 0$. Even though the optimal cost after such a transformation will be different, it will be predictably scaled by $\lambda$. We wish to regard such scaled problems as equivalent.
\end{enumerate}

Next, we describe some transformations that make important changes to instances. For every simplifying transformation that can be applied to a Set Cover problem there is an analogous transformation for Set Packing problems.

\begin{enumerate}[(a)]
\setcounter{enumi}{3}
\item 
For a Set Cover instance, if any set has a cost that is non-positive, we can take that set and remove all of its elements from the other input sets. Since it costs nothing or less than nothing to include, there must be some optimal cover that includes it. For a Set Packing instance, any set with non-positive weight can be rejected immediately, since it does not increase the total weight and its inclusion may hurt our ability to take other sets. There must be an optimal packing that excludes such a set, so we can safely exclude it.

In the literature, it is usually simply assumed that all costs/weights are positive. This transformation explains why that requirement does not diminish the generality of instances with only positive costs. Every one of the transformations discussed here will allow us to analogously trim the space of SCP and SPP instances with which we should be prepared to deal.

\item
If some input set is empty (and assuming that it has positive cost, as the above permits us to), then it will not be in any optimal cover since it incurs a cost and does no work towards our goal, so we can remove it from the problem. It will, however, be included in every optimal packing, so we can require that it is taken.

\item
Consider an instance where some universe element is not present in any of the input sets.\footnote{For the way that we have defined the problem, this is not possible, since the universe $\calU$ is defined as the union of the input sets. We describe this only for completeness.} This means that no set cover exists for the instance. It is immediately infeasible and any attempt to find a cover will fail. For Set Packing, however, we can remove the uncovered elements from the universe and work with that equivalent simpler problem instead.

\item
Consider an instance where some universe element is present in exactly 1 of the input sets. For SCP, the sole input set including this element must be included in every cover, so we can immediately include it and reduce the problem respecting that inclusion. For SPP, we can remove that element from the universe and the set that contains it. Since the constraint that every element must be included no more than once cannot fail to be satisfied for this element, we can leave it out of the problem entirely.
\end{enumerate}

We call these 4 transformations \emph{basic preprocessing}. When we talk about applying basic preprocessing to an instance, we mean that we apply all of these steps to an instance recursively until no more can be applied.

%%%%%%%%%%%%%%%%%%%%%%%%%%%%%%%%%%%%%%%%%%%%%%%
\section{Subsumption Testing}
\label{sec:Subsumption Testing}
%%%%%%%%%%%%%%%%%%%%%%%%%%%%%%%%%%%%%%%%%%%%%%%

There is an additional pair of preprocessing simplifications that we call \emph{subsumption} preprocessing, due to the similarity they share with subsumption in the context of SAT instances.
One of them operates from the perspective of the input sets, sometimes permitting the removal of an input set since in any cover/packing it can always be replaced by another input set with no loss of quality. The other operates from the perspective of the elements, enabling us to remove an element when some other element enforces a strictly stronger constraint on the problem.

\begin{enumerate}[(a)]
\setcounter{enumi}{7}
\item
If there are 2 sets $S_i,S_j$ such that $S_i\subseteq S_j$ and $c_i \geq c_j$, then we can see that $i$ need not ever be included in a cover. In any cover using $i$, we can instead replace it with $j$ to find a cover that is no more expensive. Therefore, we can remove set $i$ from the problem.

For SPP, if there are 2 sets $S_i,S_j$ such that $S_i \subseteq S_j$ and $c_i \geq c_j$, then we can remove set $j$ from the instance. Any packing using $j$ would run up against at least as many constraints as one using $i$, and it would do so for no more gain that $i$.

Basic preprocessing step (e) can be seen as a special case of this step.

\item
For any element $e \in \calU$, let $N_e = \set{i\in\calI~|~ e\in S_i}$ be the set of input sets of which $e$ is a member.

For SCP, if there are two different elements $e,e' \in \calU$ with $N_e \subseteq N_{e'}$, $e$ is a stronger constraint, and every time one of its elements is selected, so will one of $e'$'s. Thus we can omit $e'$, removing it from every set in $N_{e'}$, and have an equivalent problem. 

For SPP, if there is a pair of different elements $e,e'\in \calU$ where $N_e \subseteq N_{e'}$, $e'$ is a strictly stronger constraint. If we remove $e$ from all sets in $N_e$, we are left with a problem with precisely the same feasible packings.

Basic preprocessing step (f) can be seen as a special case of this step.

\end{enumerate}

Running subsumption testing has high complexity relative to the basic preprocessing steps. With the most na\"ive 
approach we need to check $\bigo{m^2}$ different cases for set subsumption and $\bigo{n^2}$ cases for element subsumption. In the algorithm we use for our experiments, the only shortcut we use is checking sets/elements whose neighbourhoods have shrunk since we last checked whether they could be involved in some subsumption.

%%%%%%%%%%%%%%%%%%%%%%%%%%%%%%%%%%%%%%%%%%%%%%%
\section{Independent Subproblem Separation}
%%%%%%%%%%%%%%%%%%%%%%%%%%%%%%%%%%%%%%%%%%%%%%%

\begin{figure}[!ht]
\centering
\includegraphics[width=0.4\textwidth]{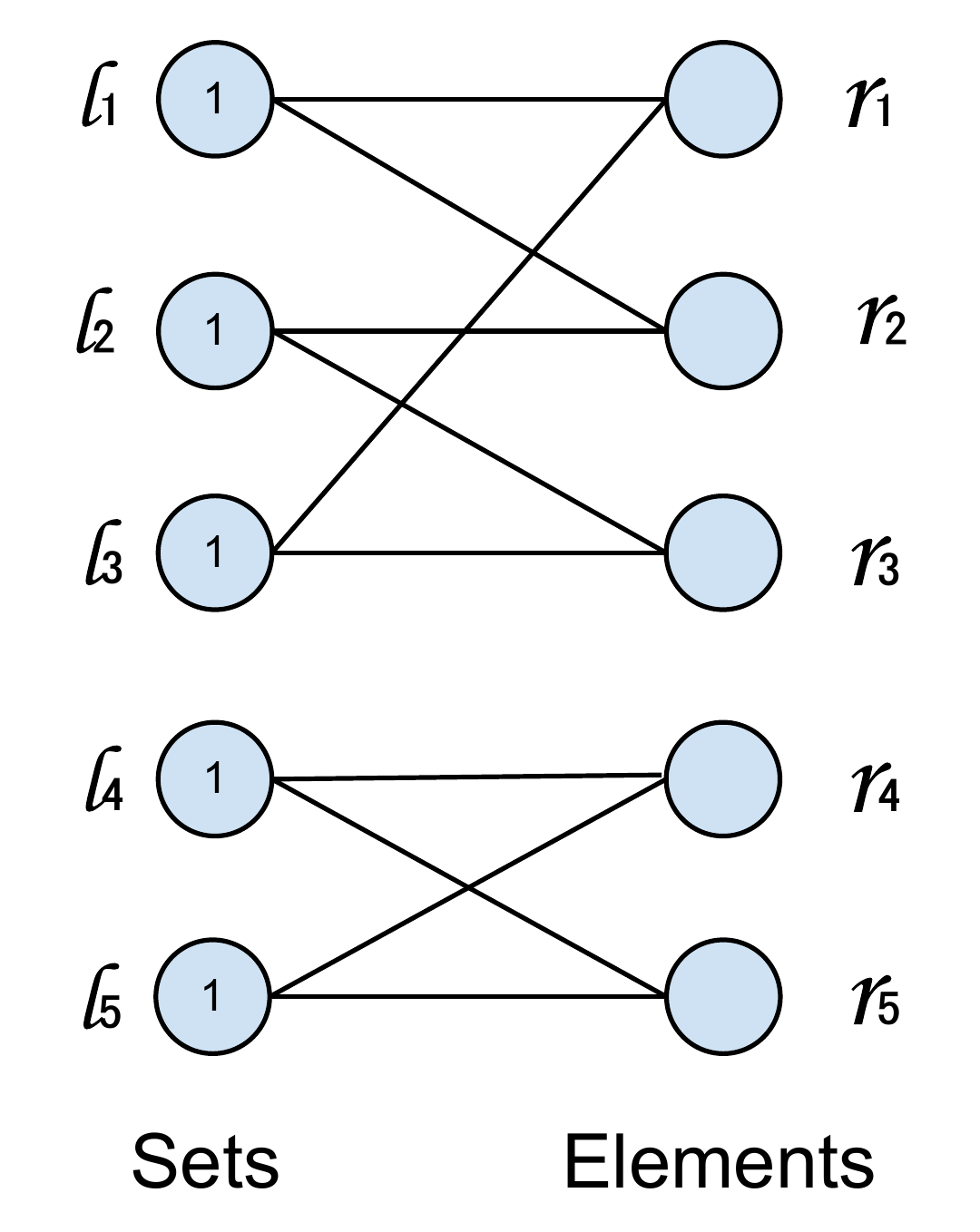}
\caption{HSC graph structure for a problem with 2 independent subproblems.}
\label{fig:independent-subproblems}
\end{figure}

Consider the set cover instance $\calI = \set{1,2,3,4,5}$ with $S_1=\set{1,2}$, $S_2=\set{2,3}$, $S_3=\set{1,3}$, $S_4=\set{4,5}$, $S_5=\set{4,5}$ and uniform costs. In figure \ref{fig:independent-subproblems} we see the HSC instance associated with this problem.
We can easily observe that choosing input sets $S_1$, $S_2$ and $S_3$ can never obtain elements 4 and 5. Likewise, sets $S_4$ and $S_5$ can never obtain elements 1, 2, or 3. We can solve instances with $\calI'=\set{1,2,3}$ and $\calI''=\set{4,5}$ independently and combine the covers for these to make a cover for the original instance. In general, we can solve each component of an HSC graph independently and combine them afterwards to obtain a cover.

Stated with reference only to the input of SCP or SPP instances, we can say that if any subset of the input sets $\calI' \subset \calI$ has the property that 
\[
\left( \bigcup_{i\in \calI'} S_i \right) \cap \left( \bigcup_{i\in \calI - \calI'} S_i \right) = \emptyset
\]
then we can solve the instances with sets $\calI'$ and $\calI - \calI'$ independently.

The basic preprocessing step (e) can be viewed as a case of this preprocessing technique.

%%%%%%%%%%%%%%%%%%%%%%%%%%%%%%%%%%%%%%%%%%%%%%%
\section{Inferring Stronger Packing Constraints}
%%%%%%%%%%%%%%%%%%%%%%%%%%%%%%%%%%%%%%%%%%%%%%%

Recalling section \ref{sec:Set Packing's Relationship to Maximum Independent Set}, a Set Packing instance corresponds to an WMIS instance on the graph $\str{\calI, E}$ where $E$ has a clique among the neighbourhood $N_e$ of every element $e\in\calU$. By finding larger cliques in this graph, we can rewrite the element neighbourhoods in the SCP instance, which may allow subsumption preprocessing to further simplify the instance.

\begin{figure}[!ht]
\centering
\includegraphics[width=1.0\textwidth]{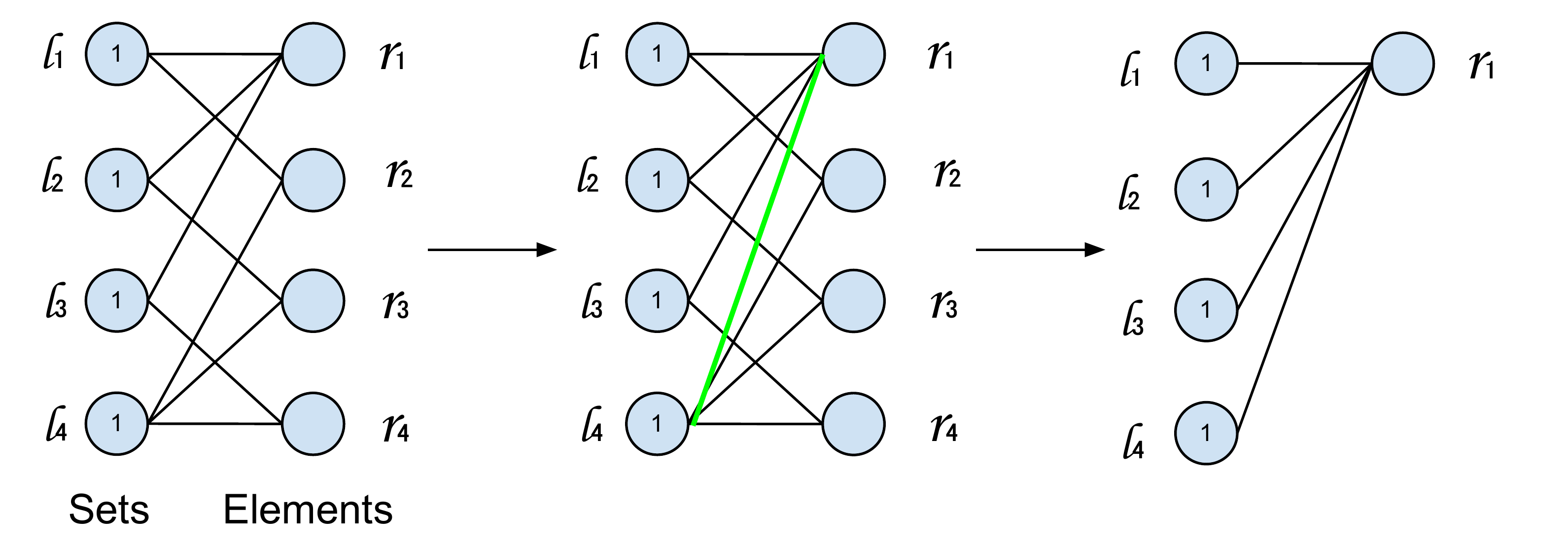}
\caption{Strengthening packing constraints. In the first step, we can see that $l_4$ intersects $l_1$ on $r_2$, $l_2$ on $r_3$, and $l_3$ on $r_4$. This permits us to add $r_1$ to $l_4$, shown as a new green edge in the central diagram. In the rightmost diagram, we have removed the other 3 elements from the instance because they are subsumed by element $r_1$.}
\label{fig:strengthening-packing-constraints}
\end{figure}

In particular, if there is some $e\in\calU$ and $i\in\calI- N_e$ for which $S_i$ intersects each of the sets in $N_e$, we can add $e$ to $S_i$ and the resulting problem will have logically equivalent constraints. We can see this process in action in figure \ref{fig:strengthening-packing-constraints}.

We can use this observation to build maximal constraints, corresponding to maximal cliques in the WMIS graph, in polynomial time. There may be multiple descriptions of the graph as maximal cliques, and we may not find the \emph{maximum} clique in our search, given the NP-hardness of the Max Clique problem. It is also not clear that this preprocessing step is always to our advantage in the context of $k$-Set Packing, since we allow ourselves to add elements to sets. Note, however, that in figure \ref{fig:strengthening-packing-constraints} we use constraint strengthening and subsumption to transform a 3-Set Packing instance into a 1-Set Packing instance, showing that this step is of use for at least some instances.

%%%%%%%%%%%%%%%%%%%%%%%%%%%%%%%%%%%%%%%%%%%%%%%
\section{Non-Minimal Covers}
%%%%%%%%%%%%%%%%%%%%%%%%%%%%%%%%%%%%%%%%%%%%%%%

In some cases the standard greedy heuristic for SCP will yield solutions that are not minimal. Let $\calH$ be the returned solution having $\bigcup_{i\in\calH} S_i=\calU$. It sometimes happens that there is a strict subset of $\calH$ that is a set cover. Consider the following instance:
\example{
\begin{align*}
\calI &= \set{1,2,3,4} \\
S_1 &= \set{1,3} \text{ with cost } c_1 = 2\\
S_2 &= \set{2,4} \text{ with cost } c_2 = 2\\
S_3 &= \set{3,4} \text{ with cost } c_3 = 1\\
S_4 &= \set{1,2} \text{ with cost } c_4 = 5\\
\end{align*}
For this instance, the initial valuations are $(1,1,2,0.4)^T$, so $S_3$ is selected. In the next iteration, the valuations are $(0.5,0.5,0.4)^T$, so $S_1$ or $S_2$ is selected with the other one joining it in the following iteration. Thus the returned set cover is $\calH=\set{S_1,S_2,S_3}$. $\set{S_1,S_2}$ is strictly smaller and is also the optimal cover.
}
In fact, it is quite common for solutions returned by the standard greedy algorithm to be redundant, so  when we want to find approximate solutions to Set Cover instances, it is often advantageous to verify that the returned solution is minimal or reduce it to a minimal solution. 
We now describe some of the techniques that have been used to reduce non-minimal approximate solutions to minimal ones.

\subsubsection{Wool and Grossman}
\label{Wool and Grossman}
In \cite{grossman1997computational}, the following technique is used to minimize approximate Set Cover solutions.

Let the Set Cover problem of interest be given by $\matA,\vecc$ as in the IP formulation, with the non-minimal solution given by the binary vector $\vecx$. Let $\vecr=\matA\vecx - \vecone$. $\vecr_i$ is the number of times that element $i$ has been redundantly acquired by the chosen sets. Then, for each set $1 \leq i \leq m$, define the minimal redundancy $\vect_i = \min_{j\in S_i} \vecr_j$. For any redundant input set $i$, $\vect_i>0$ and it can be eliminated from our cover. We choose the set $b$ for which $\vect_b$ is largest (breaking ties by choosing the largest cost set), eliminate it from our cover by setting $\vecx_b=0$ and continue until $\vect$ has 0's in every component.

%A variation on this might be to select the set $b$ for which $\vecc_b \vect_b$ is largest, taking the costs of the elements into account.

%\subsubsection{Drop Most Recent}
%Given a non-minimal solution $\calH$ returned by our heuristic algorithm, we can examine the selected sets in order from the most recently chosen to least recently chosen. If the set is redundant, we will drop it from our solution and continue. 
%The intuitive justification for this process is that sets chosen earlier were more desirable at the time they were selected, and so should be spared consideration until later.

\subsubsection{Recursive Solution}
It is possible to formulate the problem of minimizing a redundant Cover as an instance of Set Cover. Given a problem specified by $\matA, \vecc$ and a redundant solution given by the binary vector $\vecx$ we can succinctly specify the IP with variable vector $\vecz \in \set{0,1}^m$:
\begin{align*}
\text{Minimize }  &\vecc^T \vecz \\
\text{subject to }&\matA \vecz \geq \vecone \\
\text{and }& \vecz \leq \vecx
\end{align*}
The constraint that $\vecz \leq \vecx$ represents the requirement that we are looking for a subset of the sets chosen in the first run through our problem.
Once we remove the columns of $\matA$ for which $\vecx_i=0$ and reduce the length of our variable vector $\vecz$ appropriately, we have an IP with the precise form of a Set Cover problem. We can then solve this problem using the same greedy process with which we obtained $\vecx$ and recursively minimize that solution. However, in order for us to guarantee that the recursive process terminates, we need to show that the resulting instance is strictly smaller than the original instance we set out to solve in the first place. This can only fail when $\vecx=\vecone$, when our greedy process selects every input set but the produced cover is redundant. 

Since our algorithm stops after every element has been covered, there is some element that was covered for the first time in the last iteration of the greedy selection process. If we first reduce the instance $\matA,\vecc$ by requiring the inclusion of the set selected in the last iteration of the previous round, we will be solving a strictly smaller instance. For instance, in the example given earlier in this section, we run the greedy algorithm 3 times on subinstances with $(\calI, \calU)$ given by, successively,  $(\set{1,2,3,4},\set{1,2,3,4})$, $(\set{2,3},\set{2,4})$, $(\set{3},\emptyset)$, resulting in the optimal cover being selected, since the final instance is trivial.

Alternatively, we could require \emph{all} sets in the redundant solution that contain any element that was chosen exactly once to be included before forming the recursive subinstance or simply doing basic preprocessing between greedy iterations. Either would ensure that the recursive instance is smaller and that our process for minimization would terminate.

\chapter{New Greedy Heuristics for Set Cover and Set Packing}
\label{chap:New Greedy Heuristics for Set Cover and Set Packing}

%%%%%%%%%%%%%%%%%%%%%%%%%%%%%%%%%%%%%%%%%%%%%%%
\section{Motivation}
%%%%%%%%%%%%%%%%%%%%%%%%%%%%%%%%%%%%%%%%%%%%%%%
In the general greedy scheme for Set Cover we have a variety of options available to us. To distinguish between them we generate a valuation for each alternative. We require that the valuations all be positive, corresponding to the fact that any choice makes \emph{some} progress towards our goal. While constructing a cover, we know that every element must be obtained at some step. Every element forms a hard constraint on our eventual solution. 

The standard greedy heuristic is to select the option which satisfies the most constraints for the least cost. If, instead, we consider each constraint not to have identical difficulty we might prefer to make the choice such that the sum of the difficulties overcome is largest per unit cost. How then, do we assign difficulties to the constraints? The standard heuristic can be seen to derive from the decision that all constraints are equally difficult to satisfy. We might instead try to evaluate the difficulty of the different constraints by referring recursively to the valuations of the choices that would satisfy them. A constraint that is satisfied by many available options with high valuations is going to be effectively less difficult to satisfy than one satisfied only by few with low valuations. Other choices are possible, but for our immediate purposes we assign constraint difficulties inversely proportional to the sum of the valuations of choices that would satisfy them.

If we write the vector of valuations as $\vecv$ and the difficulties as $\vecd$, we have
\begin{align*}
\vecv &= \matC^{-1} \matA^T \vecd \\
\vecd &= (\matA \vecv)^\hinv
\end{align*}
This gives us a recursive definition of the valuation that we originally sought to find. In the following we show that this pair of definitions (and some variations) are satisfied only by a unique vector of valuations $\vecv$, can be found fairly quickly, and yield results that are generally preferable to those yielded by the standard heuristic.

We are interested in this general idea, generating the valuations for our current options by defining the valuation recursively, and think that it can be effectively applied in a variety of circumstances. We also develop an intuitive heuristic for Set Packing using this idea and show that it performs better than other simple greedy heuristics. We are unable to prove anything about the quality of the covers and packings obtained from our heuristics, beyond our experimental results, but hope that future work may be able to whether they yield some worst case approximation guarantee. 

More importantly, we believe that the particular heuristics we have defined can be of practical use immediately and that the general idea of calculating valuations recursively can be used effectively for a wide variety of tasks.

%%%%%%%%%%%%%%%%%%%%%%%%%%%%%%%%%%%%%%%%%%%%%%%
\section{The New Greedy Set Cover Heuristic}
%%%%%%%%%%%%%%%%%%%%%%%%%%%%%%%%%%%%%%%%%%%%%%%

%%%%%%%%%%%%%%%%%%%%%%%%%%%%%%%%%%%%%%%%%%%%%%%
\subsection{Relationship to the Standard Heuristic}
%%%%%%%%%%%%%%%%%%%%%%%%%%%%%%%%%%%%%%%%%%%%%%%

The standard greedy heuristic works by assigning a valuation for the set $i$ of $\vecv_i=|S_i| / \vecc_i$  (that is, the number of input sets hit per unit cost) and irrevocably including the set maximizing this quantity to a set cover that we are building. We then reduce the instance in accordance with the newly obtained elements no longer constraining our future decisions, redo the preprocessing and continue on until we have hit every element.

Consider the standard greedy heuristic as our starting point
\[
\vecv^{(1)} = \matC^{-1} \matA^T \vecone
\]
Given these valuations, we can consider how \emph{difficult} it might be to hit particular universe elements. Since we are less likely to select sets with low valuations, we might assign the inverse of the sum of the valuations of the sets of which it is a member. That is:
\[
\vecd^{(1)} = \left( \matA \vecv^{(1)} \right)^\hinv
\]
From this, we might wish to continue the process, letting the valuations of sets be the cost-weighted sum of the difficulties of their elements and recompute the difficulties, so we define:
\begin{align*}
\vecv^{(i)} &= \matC^{-1} \matA^T \vecd^{(i-1)} \\
\vecd^{(i)} &= \left( \matA \vecv^{(i)} \right)^\hinv
\end{align*}
We can now calculate $\vecv^{(i)}$ for any $i$ we like. In practice, as we use $\vecv^{(i)}$ for larger $i$ as a valuation for greedy set cover approximation, the resulting covers tend to improve.

We can instead consider what valuation $\vecv \in \R^m_+$ and difficulty $\vecd \in \R^n_+$ could be selected so that simultaneously $\vecv = \matC^{-1} \matA^T \vecd$ and $\vecd = \left( \matA \vecv \right)^\hinv$. 
For such a pair of vectors, we would have 
\[
\vecd = \left( \matA \matC^{-1} \matA^T \vecd \right)^\hinv
\]

%%%%%%%%%%%%%%%%%%%%%%%%%%%%%%%%%%%%%%%%%%%%%%%
\subsection{Consistent Valuations}
%%%%%%%%%%%%%%%%%%%%%%%%%%%%%%%%%%%%%%%%%%%%%%%
In general, we can write a heuristic by recursively defining two vectors $\vecv \in \R_+^m$ and $\vecd \in \R_+^m$ and finding some pair of vectors that satisfy the definition. When there is some pair of vectors satisfying this recursive definition, we will call them a \emph{consistent valuation} and may use $\vecv$ as our greedy heuristic. We require that these vectors be strictly positive because we find the notion of a negative value or difficulty incoherent in this setting, since regardless of how poor a choice some set may be, it must make some positive progress towards the goal of collecting every basis element. The choice of 
\begin{align*}
\vecv &= \matC^{-1} \matA^T \vecd \\
\vecd &= \vecone
\end{align*}
yields the standard heuristic. For a recursive definition of this sort we would hope that a consistent valuation exists and would ideally be unique. For the definitions immediately above it is clear that, for any particular instance, there is a unique consistent valuation $\vecv, \vecd$ satisfying it.

%%%%%%%%%%%%%%%%%%%%%%%%%%%%%%%%%%%%%%%%%%%%%%%
\subsection{A Family of Heuristics}
%%%%%%%%%%%%%%%%%%%%%%%%%%%%%%%%%%%%%%%%%%%%%%%

The form that we wish to propose is
\begin{align*}
\vecv &= \matC^{-1} \matA^T \vecd \\
\vecd &= \left( \matA \matC^\gamma \vecv \right)^\hinv
\end{align*}
where $\gamma \in \R$ is a free parameter that we will fix only later. For now we consider all valuations generated by $\gamma$ ranging over the reals. We find that $\gamma=-3$ performs well.$\gamma$ can be viewed as additionally penalizing sets with high cost.

If we can find a $\vecd$ for which 
\begin{align*}
\vecd
&= \left( \matA \matC^\gamma \vecv \right)^\hinv \\
&= \left( \matA \matC^{\gamma-1} \matA^T \vecd \right)^\hinv
\end{align*}
then we have a consistent valuation, since we can immediately calculate $\vecv$ given $\vecd$, and its value is uniquely determined by $\vecd$. If we have additionally determined that there is a unique $\vecd$ for which the above holds, then the consistent valuation given by our recursive definition is also unique.

Let us write $\matM =  \matA \matC^{\gamma-1} \matA^T$. $\matM$ is symmetric positive semi-definite, since 
\[
\matM = (\matC^{\frac{\gamma-1}{2}}) \matA^T)^T ((\matC^{\frac{\gamma-1}{2}}) \matA^T)
\]
If we consider only instances with no empty input sets, which we can do by requiring that basic preprocessing be done on instances, every diagonal element of $\matM$ is strictly positive. Also, if we consider only instances for which every set's cost is strictly positive, which can be accomplished again by doing basic preprocessing, every component of $\matM$ is non-negative.
From the results of chapter \ref{chap:Mathematical Results}, these qualities ensure that there is a unique $\vecd$ satisfying $\vecd = \left( \matM \vecd \right)^\hinv$, and thus that our recursive definitions specify a unique consistent valuation for every Set Cover instance that has undergone basic preprocessing.

%%%%%%%%%%%%%%%%%%%%%%%%%%%%%%%%%%%%%%%%%%%%%%%
\subsection{Relationship to Theory}
%%%%%%%%%%%%%%%%%%%%%%%%%%%%%%%%%%%%%%%%%%%%%%%

In our experiments, we will be most concerned with the heuristic obtained by setting the value $\gamma=-3$. It can be argued that $\gamma=0$ is more natural, defining the difficulty of an element as the reciprocal of the sum of valuations of the sets it is in, instead of having the terms of this sum weighted by the inverse cube of the set's cost, but we have found that using $\gamma=-3$ performs better than other values of $\gamma$ in an average case sense. An experiment justifying this choice can be seen in section \ref{sec:Varying gamma for the New Heuristic}. We have not been able to determine, for any value of $\gamma$, whether the new cover heuristic has some worst-case guaranteed approximation ratio.

Assuming that valuations given by the new cover heuristic can be calculated or approximated arbitrarily well in polynomial time (for which we provide evidence in section \ref{sec:Running Time of the New Set Cover Heuristic}) and some plausible complexity assumptions, Feige's result \cite{DBLP:journals/jacm/Feige98} shows that there must be classes of Set Cover instances for which the approximation ratio obtained using the new heuristic exceeds $\ln(n) - c\ln(\ln(n))^2$ for some $c>0$ for sufficiently large $n$. We have been unable to find any class of instances for which the new heuristic gives approximation ratios exceeding a constant as $n$ grows arbitrarily large, though we expect that such classes of instances must exist. Feige's work could theoretically be used to build explicit instances, though it would be highly inconvenient to do so.

%\workingnote{
%It would be very inconvenient. Feige's proof goes through constructions using multi-prover proof systems, and I can't claim to have anything like a good understanding of it.
%
%This might be a good place to describe the construction I have related to the hard instances for standard greedy. Of course, it doesn't actually yield hard instances, petering out with approximation ratio approaching a constant while $n$ grows. Perhaps just mentioning that the standard greedy hard example is solved exactly by this heuristic is enough for here. The construction that doesn't make hard instances is probably a very weak sort of evidence that this heuristic does well, and perhaps I only want to include it in this document because it took me some time and I thought it was kinda clever.
%}

%In section {\undone}, we show a construction related to the class of instances for which the standard heuristic gives $H(n)$ approximation ratios for which the new heuristic performs relatively poorly, but for which, as $n$ increases, the obtained approximation ratio is bounded above by a constant. 
%We expect that classes of instances with poorer approximation ratios must exist, but have been unable to discover any.

%%%%%%%%%%%%%%%%%%%%%%%%%%%%%%%%%%%%%%%%%%%%%%%
\section{The New Greedy Set Packing Heuristic}
\label{sec:The New Greedy Set Packing Heuristic}
%%%%%%%%%%%%%%%%%%%%%%%%%%%%%%%%%%%%%%%%%%%%%%%

A common greedy heuristic for WMIS is to select the vertex with largest weight divided by neighbourhood size. Considering the relationship between WMIS and SPP, this heuristic can be transferred to work on Set Packing instances. For every input set $i\in \calI$, the valuation of $i$, $v_i = c_i / (|\set{j\in\calI ~|~ S_j \cap S_i \neq \emptyset}|-1)$ the set's weight per intersecting input set besides itself. The $-1$ is to exclude the input set from being counted among its neighbours. In the unweighted case, this heuristic behaves identically whether we include or exclude the set as a neighbour of itself. We could instead consider each set a neighbour of itself, but experimentally both valuations achieve good quality packings so we will focus on the WMIS heuristic, which does not consider a set to be in its own neighbourhood.
We call this the valuation the \emph{MIS heuristic} for Set Packing.

For the IP representation of SPP, we can write the valuation vector for the standard heuristic very simply. Define the \emph{binarization} of a matrix $bin : \R^{n\times n} \to \set{0,1}^{n\times n}$ by letting $bin(\matM)$ be the matrix having zeros where $\matM$ has zeros, and 1's where $\matM$ has non-zeros. Thus 
\[
bin(\matM)_{i,j} = \left\{
     \begin{array}{lr}
       0 &  \text{if }\matM_{i,j} = 0 \\
       1 &  \text{if }\matM_{i,j} \neq 0
     \end{array}
   \right.
\]
With this function, the standard heuristic's valuation is written $\vecv = \matC ((bin(\matA^T \matA)-\matI)\vecone)^\hinv$. Note that $bin(\matA^T \matA)\vecone - \matI$ is precisely the adjacency matrix of the WMIS instance equivalent to the SPP instance with constraints given by $\matA$.

Our valuation comes from trying to generalize this idea. Instead of valuing a set as its weight divided by the quantity of neighbours it has, we could value it as its weight divided by the sum of its neighbours' weights, or its weight divided by the sum of its neighbours' ratios \emph{weight divided by neighbourhood size} and so on. This leads us to the following recursive definition for our new valuation:
\[
\vecv = \matC (bin(\matA^T \matA) \vecv)^\hinv
\]
Letting $\matM = \matC^{-1} ~bin(\matA^T \matA)$, we have $\vecv = \matM\vecv^\hinv$. If we insist on basic preprocessing, $bin(\matA^T \matA)$ will have 1's on its diagonal and no negative entry. Left multiplying this by $\matC^{-1}$, which is diagonal with strictly positive diagonal entries, results in a matrix $\matM$ with no negative entries and all diagonal entries strictly positive. From the result of section \ref{sec:Fixed Point Existence}, we know that there must be some strictly positive vector $\vecv$ satisfying our recursive definition. In general, it is not the case that $\matM$ will be positive semi-definite, so we cannot assert that it is unique. We will discuss the question of the uniqueness of our new Set Packing valuation in some detail in section \ref{sec:Fixed Points For Broader Classes of matM}. The results of our experiments with random instances, comparing the weight of packings produced by the new and 3 other heuristics, are presented in section \ref{sec:Set Packing Results}.

We have tried to insert a parameter analogous to the new cover valuation's $\gamma$, but for all variations we have attempted, the straightforward definition above appears to have performance at least good as similar alternatives.

%%%%%%%%%%%%%%%%%%%%%%%%%%%%%%%%%%%%%%%%%%%%%%%
\chapter{Mathematical Results}
\label{chap:Mathematical Results}
%%%%%%%%%%%%%%%%%%%%%%%%%%%%%%%%%%%%%%%%%%%%%%%

The aim of this section is to investigate what conditions (on a $n\times n$ matrix $\matM$) are sufficient for the existence and uniqueness of \emph{positive} solutions $\vecv\in\R^n_+$ for systems of equations of the form
\[
(\matM \vecv)^\hinv = \vecv
\]
As far as we are aware, it is not known how to characterize the $\matM$ for which there is a unique positive fixed point of this system. We have been particularly interested in $\matM$ that can be generated by our Set Cover and Packing heuristics, but there seem to be some $\matM$ that do not satisfy our assumptions which nevertheless have a unique positive solution.

For convenience, we define the function
\[
F(\vecv) = (\matM \vecv)^\hinv
\]
Any fixed point of $F$ is a solution to our system.

We will use the following assumptions about our matrix $\matM$ here. The first two are sufficient to prove the existence of a positive fixed point of $F$, while existence and (c) enable us to show that the positive fixed point is unique. In what follows, we will refer to these from time to time.
\begin{enumerate}[(a)]
\item
	For all $1 \leq i \leq n$ and $1\leq j\leq n$, $\matM_{i,j} \geq0$. Every matrix entry is non-negative.
\item
	For all $1\leq i\leq n$ we have  $ \matM_{i,i}> 0$. That is, every diagonal entry in the matrix is strictly positive.
\item
	$\matM$ is positive semi-definite. For every vector $\vecv \in \R^n$, $\vecv^T \matM \vecv \geq 0$. 
\end{enumerate}

%%%%%%%%%%%%%%%%%%%%%%%%%%%%%%%%%%%%%%%%%%%%%%%
\section{Fixed Point Existence}
\label{sec:Fixed Point Existence}
%%%%%%%%%%%%%%%%%%%%%%%%%%%%%%%%%%%%%%%%%%%%%%%

To prove the existence of a fixed point, we will show that the function $G$ has a zero. 
\[
G(\vecd) = \matM (\vecd^\hinv) - \vecd
\]
This will be accomplished by an application of the Poincar\'e-Miranda theorem (a generalization of the intermediate value theorem), as described in Idczak and Majewski\cite{poincare}.
A statement of this theorem is the following: 
%Let $\mathcal P = [\alpha,\beta]^n$ 
Let $\mathcal P = [\alpha_1,\beta_1]\times\ldots\times[\alpha_n,\beta_n]$ 
%with $\alpha \leq \beta$ set to fixed real constants, 
with $\alpha_i \leq \beta_i$ set to fixed real constants for all $1\leq i\leq n$, 
and $G: \R^n\to \R^n$ be a function that is continuous over $\mathcal P$.
Then, if the following two statements hold,
\begin{align*}
&G_i(\vecd)\geq 0 \text{ for every } \vecd\in \mathcal P \text{ for which } \vecd_i=\alpha_i \\
\text{and } &G_i(\vecd)\leq 0 \text{ for every } \vecd\in\mathcal P \text{ for which } \vecd_i=\beta_i
\end{align*}
there must be some $\vecd\in\mathcal P$ for which $G(\vecd)=\veczero$.

By definition, $G(\vecd)$'s components for $1\leq i \leq n$ are given by
\[
G_i(\vecd) = \sum_{j=1}^n \frac{\matM_{i,j}}{\vecd_j} - \vecd_i 
\]
Our function $G$'s argument is inverted relative to $F$'s. At the conclusion of the existence proof, we explain why a zero of $G$ implies the existence of a fixed point of $F$.

We use $G$ as defined above and now fix the $\alpha$'s and $\beta$'s. 
\begin{align*}
\alpha_i &=
\sqrt{\matM_{i,i}} \\
\beta_i &
=\sum_{j=1}^n \frac{\matM_{i,j}}{ \alpha_{j}}
\end{align*}
%$\alpha$ is the square root of the smallest diagonal entry in $\matM$ and 
Note that fact (b) guarantees that the $\alpha$'s are positive.

First, we demonstrate that $\alpha_i \leq \beta_i$, for all $1\leq i\leq n$ as required. 
\begin{align*}
\alpha_i
&= \sqrt{\matM_{i,i}} \\
&= \matM_{i,i} \alpha_i^{-1} \\
&\leq \matM_{i,i} \alpha_i^{-1} + \sum_{j\neq i} \matM_{i,j} \alpha_j^{-1}\\
&= \sum_{j = 1}^n \matM_{i,j} \alpha_j^{-1} \\
&= \beta_i
\end{align*}
For the inequality, we use fact (a) and the fact that the $\alpha_i$ are strictly positive.

Also, we can see that $G$ is continuous over $\mathcal P$. The only problem we might encounter is if $\vecd_i$ is 0 for some $i$. Since $\alpha_i>0$, $\mathcal P$ has no such points and we are therefore quite safe.

There are two things left to do.
First, assume, for arbitrary $1\leq i\leq n$, that we have $\vecd \in \mathcal P$ with $\vecd_i = \alpha_i$. We must now show that this assumption guarantees that $G_i(\vecd)\geq 0$. Since $G_i(\vecd) = \sum_{j=1}^n \frac{\matM_{i,j}}{\vecd_j} - \vecd_i $, it is sufficient to show that $\vecd_i \leq \sum_{j=1}^n \frac{\matM_{i,j}}{\vecd_j}$, which we derive below.

\begin{align*}
\vecd_i 
&=     \alpha_i \\
&=     \frac{\alpha_i^2}{\alpha_i} \\
&=     \frac{ \matM_{i,i}} {\alpha_i} \\
&=     \frac {\matM_{i,i}} {\vecd_i} \\
&\leq  \frac {\matM_{i,i}} {\vecd_i} + \sum_{j\neq i} \frac{\matM_{i,j}} {\vecd_j}\\
&= \sum_{j=1}^n \frac {\matM_{i,j} } {\vecd_j} 
\end{align*}

Second, assume, for arbitrary $1\leq i\leq n$, that we have $\vecd \in \mathcal P$ with $\vecd_i = \beta_i$. We must now show that this assumption guarantees that $G_i(\vecd)\leq 0$. This is equivalent to $\vecd_i \geq \sum_{j=1}^n \frac{\matM_{i,j}}{\vecd_j}$, which we will now show.
\begin{align*}
\vecd_i
&=    \beta_i \\
&=    \sum_{j=1}^n \frac{\matM_{i,j}} {\alpha_{j}} \\
&\geq \sum_{j=1}^n \frac{\matM_{i,j}} {\vecd_j} 
\end{align*}
These facts allow us to apply the Poincar\'e-Miranda theorem, establishing the existence of some vector $\vecd \in \mathcal P$ such that $G(\vecd) = \veczero$.  For such a vector, we can define $\vecv=\vecd^\hinv$, and show that it is a fixed point of $F$.
\begin{align*}
F(\vecv) 
&= (\matM \vecv)^\hinv \\
&= (\matM \vecd^\hinv)^\hinv \\
&= (\matM \vecd^\hinv - \vecd + \vecd)^\hinv \\
&= (G(\vecd) + \vecd)^\hinv \\
&= \vecd^\hinv \\
&= \vecv
\end{align*}
So $\vecv$ is a fixed point of $F$. All of our assumptions were stated explicitly, so we now assert that for matrices $\matM$ satisfying (a) and (b) there must be a positive vector $\vecv$ for which $(\matM\vecv)^\hinv = \vecv$.

Since there is a root of $G$ in $[\alpha_1,\beta_1]\times\ldots\times[\alpha_n,\beta_n]$, there must be a fixed point of $F$ in $[\beta_1^{-1},\alpha_1^{-1}]\times\ldots\times[\beta_n^{-1},\alpha_n^{-1}]$.
Switching this to $F$-like vectors and simplifying, we have that there exists a fixed point $\vecv$ of $F$ satisfying the following bounds:
\[
F(diag(\matM)^{\diamond-1/2})\leq \vecv \leq diag(\matM)^{\diamond-1/2}
\]
\section{Fixed Point Uniqueness}
%%%%%%%%%%%%%%%%%%%%%%%%%%%%%%%%%%%%%%%%%%%%%%%

Assume that $F$ has a fixed point. That is, there is some $\vecv \in \R_+^n$ with $F(\vecv)=(\matM\vecv)^\hinv=\vecv$. Because of this, $\matM\vecv = \vecv^\hinv$, which we will use below. We will show that the additional requirement that $\matM$ is positive semi-definite is sufficient to guarantee  that $\vecv$ is the only positive fixed point of $F$.
%We will also require that $\matM$ is positive semi-definite, (i.e. for all $\vecx\in \R^n, \vecx^T\matM\vecx \geq 0$).

Consider the function
\[
K(\vecu)=(\vecu-\vecv)^T(\matM\vecu-\vecu^\hinv)
\]
This is a scalar-valued inner product of 2 vectors. We will show that for all $\vecu\in\R_+^n$ with $\vecu\neq \vecv$, $K(\vecu) > 0$. The existence of any fixed point $\vecu \in \R^n_+$ besides $\vecv$, then, is a contradiction, since any fixed point of $F$ would have 
\begin{align*}
K(\vecu) 
&= (\vecu-\vecv)^T(\matM\vecu-\vecu^\hinv) \\
&= (\vecu-\vecv)^T(\vecu^\hinv-\vecu^\hinv) \\
&= (\vecu-\vecv)^T(\veczero) \\
&= 0
\end{align*}

%To see that this implies the fixed point's uniqueness, consider another fixed point of $F$, $\vecx$ with $vecx\neq \vecv$. Then $K(\vecx)=0$, since the rightmost term $(\matM\vecx-\vecx^\hinv)$ of the expression defining $K$ is $0$ at any fixed point of $F$. Thus the existence of any fixed point besides $\vecv$ yields a contradiction.

Let $\vecu\in\R_+^n$ with $\vecu\neq \vecv$. To establish that $K(\vecu)$ must be positive, we will start by subtracting $(\vecu-\vecv)^T\matM(\vecu-\vecv)$ and simplifying this. Note that this quantity is non-negative, since $M$ is positive semi-definite.
\begin{align*}
K(\vecu)
&=(\vecu-\vecv)^T(\matM\vecu-\vecu^\hinv) \\
&\geq (\vecu-\vecv)^T(\matM\vecu-\vecu^\hinv) - (\vecu-\vecv)^T\matM(\vecu-\vecv)\\
&= (\vecu-\vecv)^T(\matM\vecu-\vecu^\hinv - \matM(\vecu-\vecv))\\
&= (\vecu-\vecv)^T(\matM\vecu-\vecu^\hinv - \matM\vecu+\matM\vecv)\\
&= (\vecu-\vecv)^T( \matM\vecv -\vecu^\hinv )\\
&= (\vecu-\vecv)^T( \vecv^\hinv -\vecu^\hinv )
\end{align*}
Writing out this last expression as a sum, we have
\[
\sum_{i=1}^n (\vecu_i - \vecv_i) \left(\frac1 {\vecv_i} - \frac 1{\vecu_i}\right)=
\sum_{i=1}^n \frac{(\vecu_i - \vecv_i)^2}{\vecu_i \vecv_i}
\]
Each term in this series is positive unless $\vecu_i = \vecv_i$, in which case it is 0. Since $\vecu \neq \vecv$, for some $i$ $\vecu_i \neq \vecv_i$. Thus at least one term in the sum will be positive, and none are negative, so the entire sum must be positive, establishing our claim that $K(\vecu)$ is positive.

\chapter{Numerical Matters}
\label{chap:Numerical Calculation}
%%%%%%%%%%%%%%%%%%%%%%%%%%%%%%%%%%%%%%%%%%%%%%%

In order to calculate the consistent valuation for the new recursively defined valuations, we need to solve systems of the form $\matM\vecv=\vecv^\hinv$ for some $\vecv \in \R^n_+$ given a matrix $\matM \in \R^{n\times n}_{\geq 0}$. In this section we describe a technique that has worked well for us in practice, discuss the rate of convergence of our technique, address the time complexity of computing our new heuristic valuations and explore the prospects for solving these sorts of systems exactly.

%%%%%%%%%%%%%%%%%%%%%%%%%%%%%%%%%%%%%%%%%%%%%%%
\section{Calculating Fixed Points In Practice}
\label{sec:Calculating Fixed Points In Practice}
%%%%%%%%%%%%%%%%%%%%%%%%%%%%%%%%%%%%%%%%%%%%%%%

Let us consider a particular matrix for which we wish to calculate a vector $\vecv$ for which $\matM\vecv=\vecv^\hinv$,
\begin{align*}
\matM = 
\left(
\begin{array}{cccc}
 3 & 2 & 1 & 2 \\
 2 & 3 & 1 & 1 \\
 1 & 1 & 2 & 1 \\
 2 & 1 & 1 & 2 \\
\end{array}
\right)\end{align*}
A na\"ive way 
of attempting to calculate such a $\vecv$ would be to start at some initial vector, possibly $v^{(0)}=(1,1,1,1)^T$, and then iteratively calculate $\vecv^{(k+1)}=(\matM\vecv^{(k)})^\hinv$ until $||\vecv^{(k-1)}-\vecv^{(k)}||$ is sufficiently small. 

In practice, this does not work. Some values of $\vecv^{(k)}$ for different $k$ are shown below:
\begin{align*}
\vecv^{(0)}& = (1,1,1,1)^T \\
\vecv^{(1)}& = (0.125, 0.142857, 0.2, 0.166667)^T \\
\vecv^{(2)}& = (0.837488, 0.95672, 1.19829, 1.07969)^T \\
\vecv^{(999)}& = (0.128746, 0.146926, 0.188802, 0.167022)^T \\
\vecv^{(1000)}& = (0.831298, 0.948681, 1.21907, 1.07844)^T \\
\vecv^{(9999)}& = (0.128746, 0.146926, 0.188802, 0.167022)^T \\
\vecv^{(10000)}& = (0.831298, 0.948681, 1.21907, 1.07844)^T \\
\end{align*}
The vector settles into a repeating cycle of length 2, getting us no closer to a fixed point.

\begin{figure}[!ht]
\centering
\includegraphics[width=0.6\textwidth]{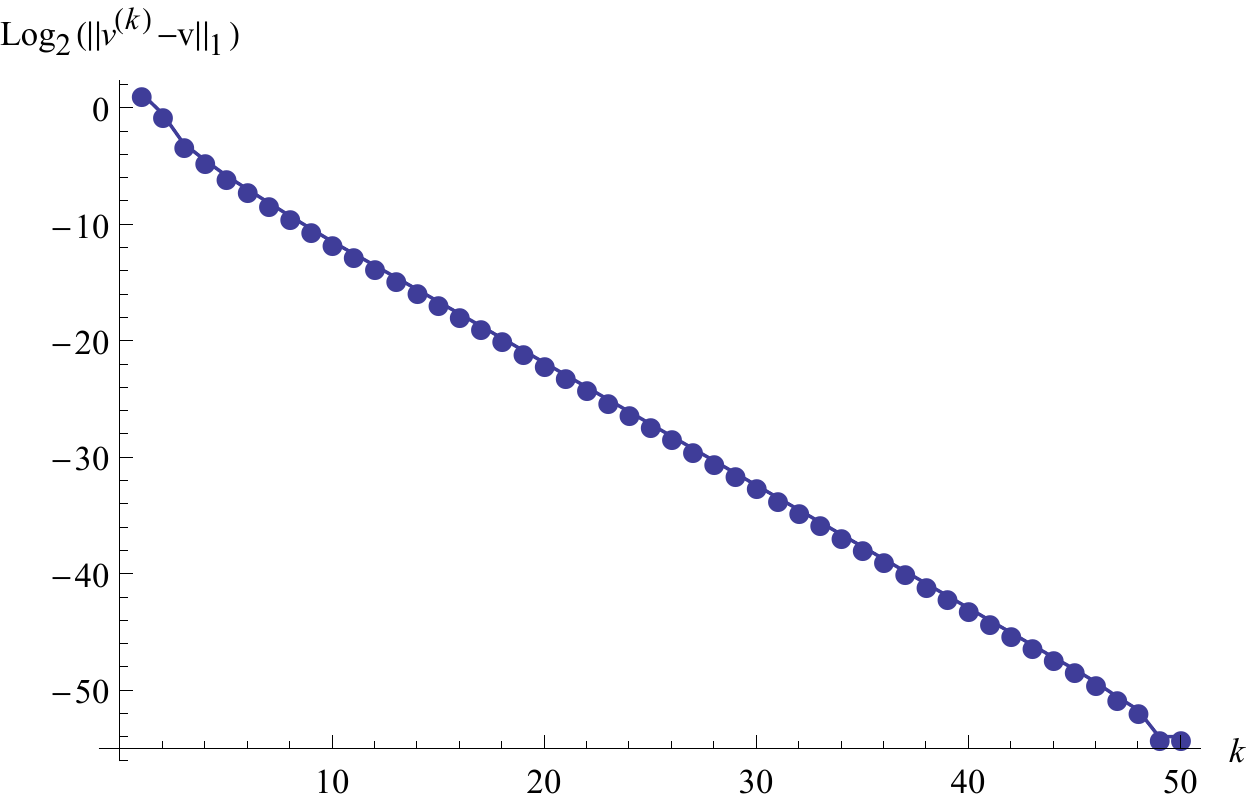}
\caption{Convergence of $\vecv^{(k)}$ towards the true fixed point $\vecv$ starting with $\vecv^{(k)}=\vecone$ and using a weighted iteration.}
\label{fig:convergence-a}
\end{figure}

We have found the following iteration to be very effective in finding fixed points. 
Let $\vecv^{(0)} \in \R^n_+$ be some positive starting vector. For $k\geq 1$, define 
\[
\vecv^{(k+1)} = \frac{ \vecv^{(k)} + (\matM \vecv^{(k)})^\hinv }{2}
\]
Every iteration is the arithmetic mean of the previous point and the na\"ive iteration's next point.
This iteration appears to have straightforward linear convergence, as can be seen in figure \ref{fig:convergence-a}, showing the distance of successive values of $\vecv^{(k)}$ from the true fixed point for this problem, $\vecv \approx (0.327149, 0.373344, 0.479752, 0.424410)^T$. Note that at the right side, convergence halts because we have obtained the fixed point to floating-point double-precision. The value that we are using for the \emph{true fixed point} is a vector computed to agree with the fixed point up to 80 decimal places for each component.

We have found that this technique consistently achieves a linear rate of convergence, regardless of the initial point used, so long as it is strictly positive. We have found no non-negative $\matM$ with strictly positive diagonals for which this iteration does not converge, though we have no proof that it must converge in all such cases.

Some simple variations of this iteration are possible and can have a significant impact on the rate of convergence. Choosing $\vecv^{(k+1)}$ to be any weighted mean of $\vecv^{(k)}$ and $(\matM \vecv^{(k)})^\hinv$ appears to be satisfactory for convergence to occur. In practice, we have found that using the arithmetic mean and putting half as much weight on $\vecv^{(k)}$ as the new term causes the series to converge fairly quickly relative to alternatives. We have also found that using the starting point with components $\vecv^{(0)}_i = (\sum_{j=1}^n \matM_{i,j})^{-1/2}$, the inverse square roof of the row sum, works particularly well. This can also be written $\vecv^{(0)} = (\matM\vecone)^{\diamond-1/2}$.

Thus, our final recommendation for calculating solutions to these systems is the following iteration:
\begin{align*}
\vecv^{(0)} &= (\matM\vecone)^{\diamond-1/2} \\
\vecv^{(k+1)} &= \frac{ \vecv^{(k)} + 2(\matM \vecv^{(k)})^\hinv }{3}
\end{align*}

\begin{figure}[!ht]
\centering
\includegraphics[width=0.6\textwidth]{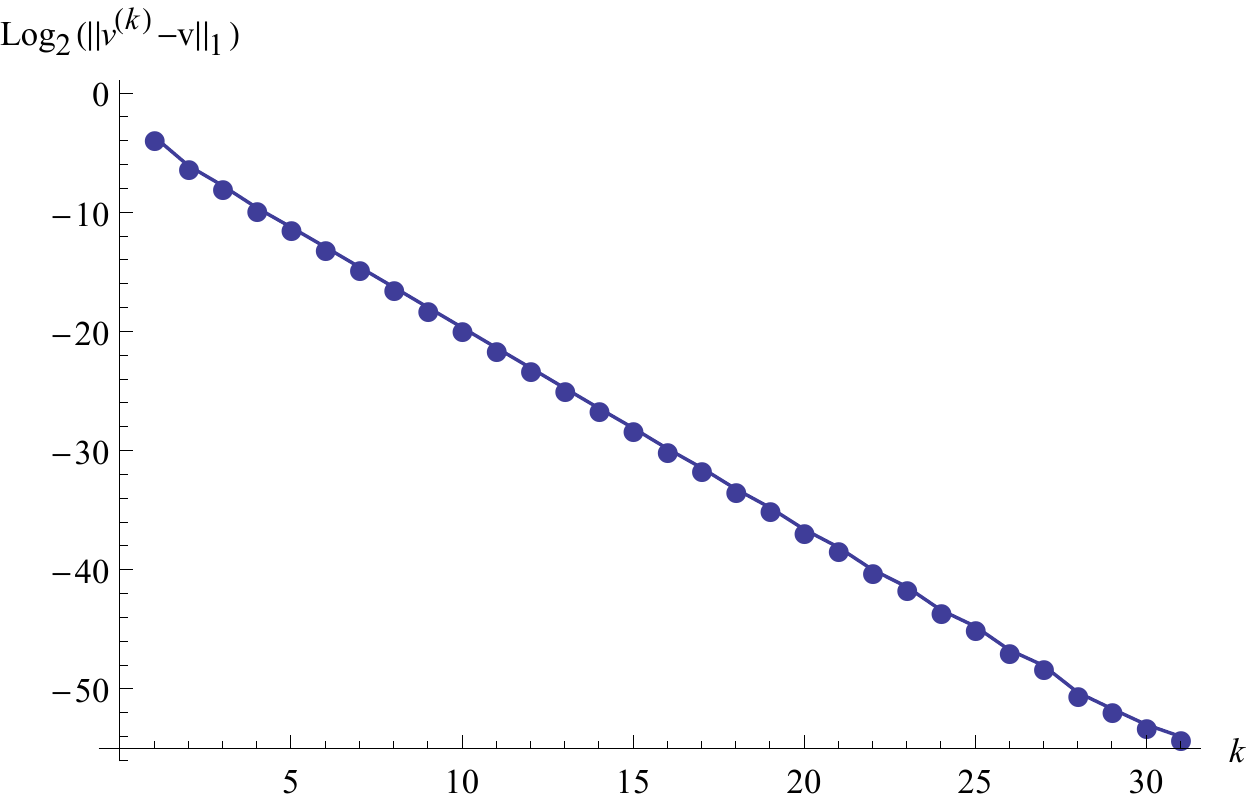}
\caption{Convergence of $\vecv^{(k)}$ for our final proposed iteration.}
\label{fig:convergence-b}
\end{figure}

Using this scheme we obtain the fixed point to floating-point double-precision in nearly 20 fewer iterations, as indicated by figure \ref{fig:convergence-b}.

\begin{figure}[!ht]
\centering
\includegraphics[width=0.6\textwidth]{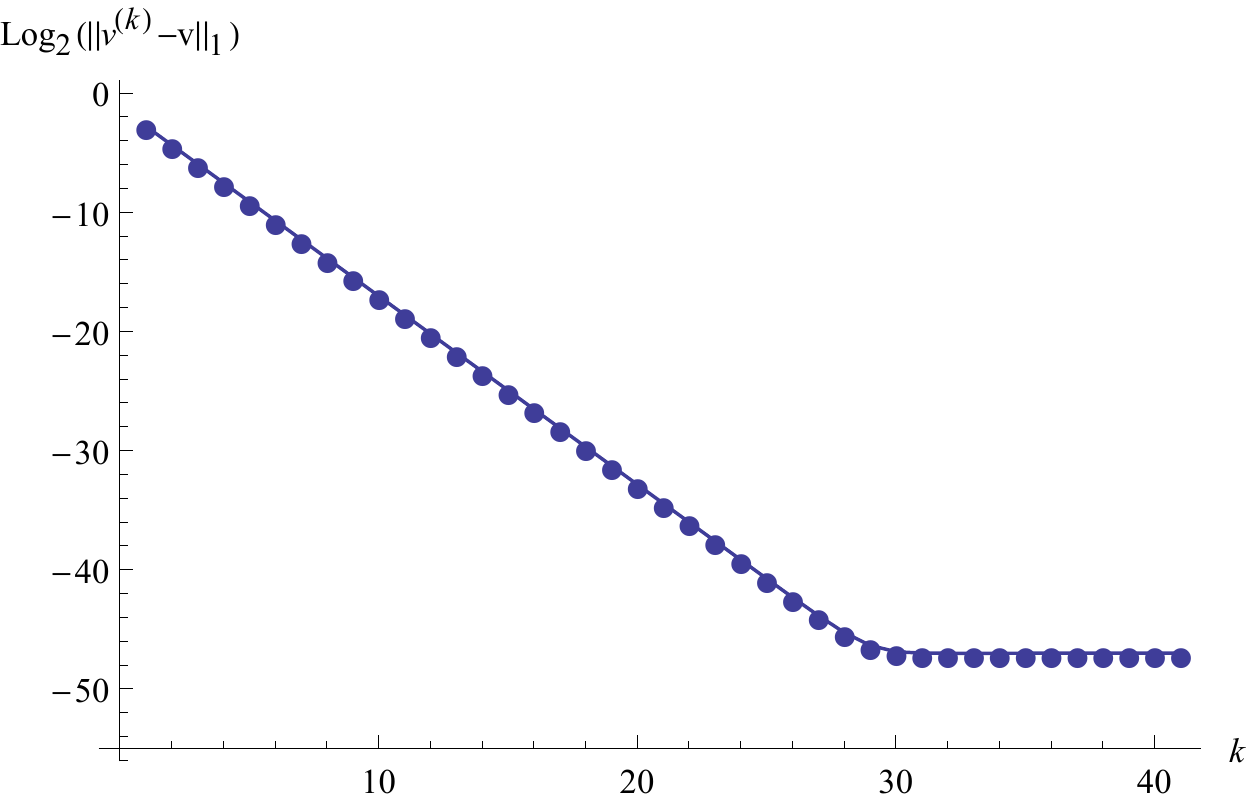}
\caption{Convergence of $\vecv^{(k)}$ for our final proposed iteration for a matrix $\matM$ of size $10000\times 10000$ with entries selected uniformly from the reals in the interval $(0,5)$}
\label{fig:convergence-c}
\end{figure}

This convergence behavior appears to be universal and is typical even for very large matrices $\matM$, as suggested by figure \ref{fig:convergence-c}. For floating-point double-precision, 30 iterations appear to suffice. It seems we can obtain the fixed point to any required precision in a number of iterations that is constant with respect to the instance size. The time required to perform one iteration is dominated by the time taken to perform the matrix multiplication. This fact makes our valuation technique competitive with the standard greedy algorithm, as we discuss further in \ref{sec:Running Time of the New Set Cover Heuristic}.

%%%%%%%%%%%%%%%%%%%%%%%%%%%%%%%%%%%%%%%%%%%%%%%
\subsection{An Alternate Iteration}
\label{sec:An Alternate Iteration}
%%%%%%%%%%%%%%%%%%%%%%%%%%%%%%%%%%%%%%%%%%%%%%%

We have found an alternative iteration scheme that is significantly different from the one presented above. It proceeds from the simple idea of asking how should $\vecv_j$ be set if all other components of $\vecv$ are correct. We want $\vecv_i = \frac 1{\sum_{j=1}^n \matM_{i,j}\vecv_j}$. Writing this as a 2nd degree polynomial in $\vecv_i$, we have 
\[
\vecv_i^2 + \frac{\left(\sum_{j\neq i} \matM_{i,j} \vecv_j\right)}{\matM_{i,i}}\vecv_i - \frac 1{\matM_{i,i}} = 0
\]
Defining $\vecs_i = \frac 1{\matM_{i,i}}\sum_{j\neq i} \matM_{i,j} \vecv_j $, the unique positive solution to this equation is
\[
\vecv_i = \frac 12 \left( \sqrt{\vecs_i^2 +4/\matM_{i,i}} - \vecs_i   \right)
\]
This leads us to propose the following iteration scheme, starting at any positive $\vecv^{(0)}$.
Let $\vecs^{(k+1)}$ be the vector with components $\vecs^{(k+1)}_i = \frac 1{\matM_{i,i}} \sum_{j\neq i} \matM_{i,j} \vecv^{(k)}_j$. This can also be found by computing $\matM \vecv^{(k)}$, dividing each component by the appropriate diagonal entry of $\matM$ and then subtracting $\vecv^{(k)}$.

Then define the new iteration $\vecv^{(k+1)}$ to be the vector with components
\[
\vecv^{(k+1)}_i = 
\frac 12 \left( \sqrt{\left(\vecs^{(k)}_i \right)^2 +4/\matM_{i,i}} - \vecs^{(k)}_i   \right)
\]

We have found that this iteration converges to a fixed point regardless of the initial point $\vecv^{(0)}$ used, although only slowly for large instances. When combined with the previous iteration in a weighted average, as is done with our previous iteration, we find that this converges to a fixed point similarly quickly.
Additionally, we have had good results with updating the valuation vector one component at a time, as we describe in section \ref{sec:Using the Alternative Iteration}.

%%%%%%%%%%%%%%%%%%%%%%%%%%%%%%%%%%%%%%%%%%%%%%%
\section{Additional Shortcuts}
%%%%%%%%%%%%%%%%%%%%%%%%%%%%%%%%%%%%%%%%%%%%%%%
We have found that we can speed this process up even more with a few  optimizations. Since we are only calculating the full valuation  in order to see which of its components is largest, very high precision is often not required and we can terminate the iteration well before reaching machine precision. At the fixed point for the random instances we have examined, we found that the difference between the valuations of the highest valuation set and the second-highest valuation set is typically around 1\%. 
Assuming that the iteration converges as straightforwardly as suggested in figure \ref{fig:convergence-c}, we should be safe halting the iteration when $||\vecv{(k)}-\vecv{(k-1)}||_1$ is less that the difference between the largest and second-largest components of $\vecv^{(k)}$. 

Note that for the new cover heuristic, this is less straightforward. The valuations are given by $\vecv = \matC^{-1}\matA^T\vecd$ where $\vecd$ is the fixed point we seek. Terminating early here should be done with the knowledge that the maximum of $\vecv$ will not change in later iterations. Conveniently, it is possible to straightforwardly examine what the valuation would be at each iteration as an intermediate result of the matrix product needed to perform the next iteration. That is, given $\vecd^{(k)}$ we can consider $\vecv^{(k)} = \matC^{-1}\matA^T\vecd^{(k)}$ and continue on to calculate $\vecd^{(k+1)} = (\matA \matC^{\gamma} \vecv^{(k)})^\hinv$.

Another shortcut we have found is reusing the valuations or difficulties found in the previous greedy iteration as a starting point for the next fixed point iteration. After the greedy scheme selects one set to include in the growing cover or packing, the instance is modified and preprocessed again, but the new system is still fairly similar to the previous one, making the use of the previous fixed point vector a better starting point than the one defined in section \ref{sec:Calculating Fixed Points In Practice}.

Additionally, we find that the performance of the overall greedy algorithm does not suffer significantly when the maximum number of iterations is fixed at even very small numbers. In random instances we have found that, after an average of around two iterations the same set is selected as would be after any number of subsequent iterations. The figure of two is slightly misleading, however, because the variance of the number of iterations before the correct choice is made is fairly large. Despite this, fixing a maximum number of iterations can be an effective way of controlling the running time of the overall algorithm without hurting its performance too badly.

%%%%%%%%%%%%%%%%%%%%%%%%%%%%%%%%%%%%%%%%%%%%%%%
\subsection{Using the Alternate Iteration for the Packing Heuristic}
\label{sec:Using the Alternative Iteration}
%%%%%%%%%%%%%%%%%%%%%%%%%%%%%%%%%%%%%%%%%%%%%%%
We have found the alternate iteration technique to be particularly useful in obtaining fixed points for the packing heuristic. Updating each component independently, we find that we reach fixed-point double precision in around 15 iterations. The form of the iterations is particularly clean for this heuristic, and we present pseudocode for the entire iteration procedure below.

\begin{algorithmic}
\ForAll{$i \in \calI$}
	\State $M_i \gets \set{j \in \calI ~|~ S_i \cap S_j \neq \emptyset} - \set{i}$ \Comment{Generate sparse version of $\matM$}
	\State $\vecv_i \gets (|M_i|/\vecc_i )^{-1/2} $ \Comment{Initialize the valuation vector}
\EndFor

\For{each iteration}
	\ForAll{$i \in \calI$}
		\State $s \gets \sum_{j\in M_i} \vecv_j$
		\State $\vecv_i \gets \frac 12 (\sqrt{s^2+4 \vecc_i} - s)$ \Comment{Update the valuation for set $i$}
	\EndFor
\EndFor
\State\Return $\vecv$
\end{algorithmic}

\begin{figure}[!ht]
\centering
\includegraphics[width=0.6\textwidth]{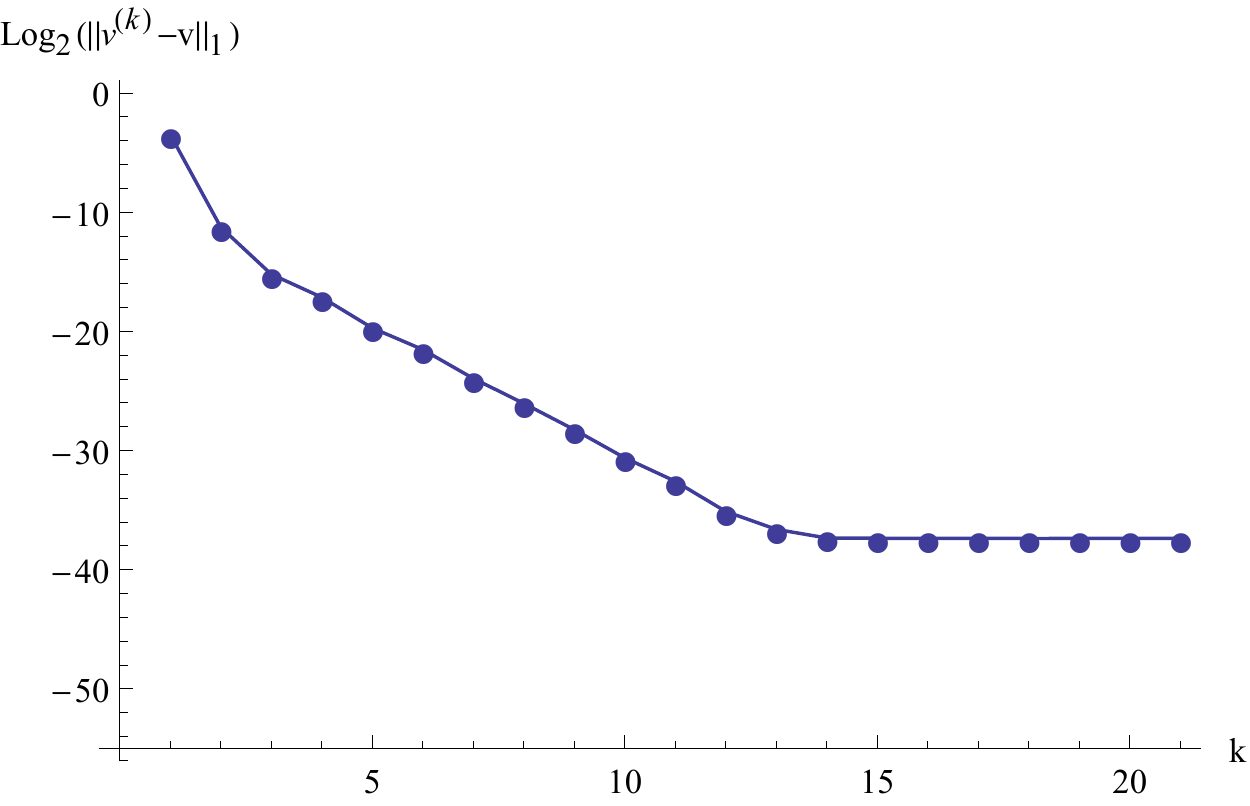}
\caption{Convergence of $\vecv^{(k)}$ towards the true fixed point for a random $5000\times 5000$ matrix.}
\label{fig:convergence of new iteration}
\end{figure}

The convergence of this approach can be seen in figure \ref{fig:convergence of new iteration}.

%%%%%%%%%%%%%%%%%%%%%%%%%%%%%%%%%%%%%%%%%%%%%%%
\section{Running Time of the New Heuristics}
%%%%%%%%%%%%%%%%%%%%%%%%%%%%%%%%%%%%%%%%%%%%%%%

%%%%%%%%%%%%%%%%%%%%%%%%%%%%%%%%%%%%%%%%%%%%%%%
\subsection{Running Time of the New Set Cover Heuristic}
\label{sec:Running Time of the New Set Cover Heuristic}
%%%%%%%%%%%%%%%%%%%%%%%%%%%%%%%%%%%%%%%%%%%%%%%

Let $\matA$ be the fundamental matrix and $\matC$ be the diagonal matrix of costs for a Set Cover instance. For many problems of practical interest and for the random instances we explore in our experiments, $\matA$ is often quite sparse. Let $\rho$ be the \emph{density} of a Set Cover instance calculated as $\rho = \frac{\sum_{i,j}\matA_{i,j}}{mn}$. In practice we use a sparse representation of $\matA$, equivalent to keeping track of every input set and the neighbourhood of every element. This enables us to perform matrix products $\matA \vecv$ and $\matA^T\vecd$ for vectors $\vecv$ and $\vecd$ in time $\bigo{mn\rho}$. Similar products using $\matC$ only take time $\bigo{m}$ since $\matC$ is diagonal.

Let $\matM=\matA \matC^{\gamma-1} \matA^T$. 
In order to calculate the new cover heuristic, we need to find $\vecd \in \R^n_+$ for which 
$\vecd = ( \matM \vecd )^\hinv$. and then calculate the valuations $\vecv = \matC^{-1}\matA^T\vecd$.
The iterative fixed point approximation technique described in section \ref{sec:Calculating Fixed Points In Practice} enables us to rapidly approximate the required $\vecd$, obtaining it to floating-point double-precision in around 30 iterations. The running time of each iteration is dominated by the time taken to perform a matrix multiplication by $\matM$. With our sparse representations of $\matA$ and $\matC$, this is again time $\bigo{mn\rho}$. This means that the time taken for one iteration of our greedy Set Cover algorithm with the new heuristic only takes a constant multiple of the time taken by the standard greedy algorithm.

%%%%%%%%%%%%%%%%%%%%%%%%%%%%%%%%%%%%%%%%%%%%%%%
\subsection{Running Time of the New Set Packing Heuristic}
%%%%%%%%%%%%%%%%%%%%%%%%%%%%%%%%%%%%%%%%%%%%%%%

The cost of approximating fixed points for our new packing heuristic is significantly worse than it is for cover. For a Set Packing instance given by $\matA$ and $\matC$, we are interested in the matrix $\matM = \matC bin(\matA^T \matA)$. In order to perform matrix multiplications by $\matM$, we are not helped substantially by the sparse data structures we maintain and our ability to do matrix multiplication by $\matA$ and $\matA^T$ efficiently.

In practice, to multiply a vector $\vecv$ by $\matM$, we find each row of $\matM$ independently and use each one to calculate a single component of $\matM\vecv$. Using the notation of input sets $S_i$ and element neighbourhoods $N_e$, we obtain $(\matM\vecv)_i$ by first finding $R_i = \bigcup_{e\in S_i} N_e$ and then calculating $(\matM\vecv)_i = \vecc_i \left(\sum_{j\in R_i } \vecv_j\right)^{-1}$. To estimate the running time of this technique, note that we need to compute the union of around $m\rho$ neighbourhoods, each of size around $n\rho$ for each of the $m$ input sets. Thus we expect the running time for one multiplication by $\matM$ to be in $\bigo{m^2n\rho^2}$, a factor of $m\rho$ larger than the time required for matrix multiplication in the case of the cover heuristic. This can make finding approximate packings with the new heuristic less practical than finding approximate covers. Note that the MIS heuristic has a similar issue, needing $\bigo{m^2n\rho^2}$ time to calculate, and it runs in around 1/30th of the time needed for the new heuristic.

%%%%%%%%%%%%%%%%%%%%%%%%%%%%%%%%%%%%%%%%%%%%%%%
\section{Exact Calculation of Fixed Points}
%%%%%%%%%%%%%%%%%%%%%%%%%%%%%%%%%%%%%%%%%%%%%%%

The principal reason that we have so few solid mathematical results here is that we find it difficult to characterize the solutions to $\matM \vecv = \vecv^\hinv$.  In a few cases, however, we can write out the solution explicitly:
\begin{enumerate}
\item
When $\matM$ is a diagonal matrix with all diagonal entries positive, there is a unique positive fixed point $\vecv$ whose components are given by $\vecv_i = \matM_{i,i}^{-\frac 12}$. In other words, $\vecv = diag(\matM)^{\diamond-1/2}$.

\item
When $\matM$ is non-negative with all non-zero diagonal entries and $\matM$'s rows all have the same sum, there is a positive fixed point with all components equal to $s^{-\frac 12}$ where $s$ is the sum of any row of $\matM$. Equivalently, there is a fixed point at $(\matM \vecone)^{\diamond-1/2}$.

\item
When $\matM$ is positive with all entries on the same row being equal, every column of $\matM$ is the same. Let $\vecu$ be one column of $\matM$, and let $s=\sqrt{\sum_{j=1}^n \frac 1 {\vecu_j}}$, the square root of the sum of the reciprocals of each row's entry. There is a positive fixed point at $\vecv = (s\vecu)^\hinv$

\item
In some other situations we can find fixed points exactly by calculating them numerically and then using an inverse symbolic calculator to find a representation of the number. For instance, when 
\[
\matM = 
\left(
\begin{array}{ccc}
 1 & 1 & 1 \\
 1 & 1 & 0 \\
 1 & 0 & 1 \\
\end{array}
\right)
\]
there must be a unique $\vecv \in \R^3_+$ satisfying $\matM\vecv=\vecv^\hinv$. Numerically, it is given by $(0.48587, 0.78615, 0.78615)^T$. It turns out that the exact value of $\vecv$ is 
\[
\left(
\begin{array}{c}
 \sqrt{\sqrt{5}-2} \\
 \sqrt{\frac{1}{2} \left(\sqrt{5}-1\right)} \\
 \sqrt{\frac{1}{2} \left(\sqrt{5}-1\right)} \\
\end{array}
\right)
\]

\end{enumerate}
In the general case we do not expect that fixed points can be described by any simple closed form. Without a deeper understanding of these fixed points, we have had to rely on experimental results to provide the force of our overall argument for the value of the new heuristics. There is much room for progress in understanding this problem.

%%%%%%%%%%%%%%%%%%%%%%%%%%%%%%%%%%%%%%%%%%%%%%%
\section{Fixed Points For Broader Classes of Matrices}
\label{sec:Fixed Points For Broader Classes of matM}
%%%%%%%%%%%%%%%%%%%%%%%%%%%%%%%%%%%%%%%%%%%%%%%

The assumptions we make about matrices $\matM$ in order to prove the existence and uniqueness of fixed points are these:
\begin{enumerate}[(a)]
\item No entries of $\matM$ are negative.
\item All diagonal entries of $\matM$ are positive.
\item $\matM$ is positive semi-definite.
\end{enumerate}
If we relax constraint (a), we can still find matrices with positive fixed points. For instance
\[
\left(
\begin{array}{ccc}
 3 & 1 & -1 \\
 1 & 3 & 1 \\
 0 & 1 & 3 \\
\end{array}
\right)
\]
appears to have a fixed point at approximately $(0.59242, 0.42200, 0.51129)^T$. 

For a similar matrix $
\left(\begin{array}{ccc}
 3 & -1 & -1 \\
 -1 & 3 & -1 \\
 0 & -1 & 3 \\
\end{array}\right)$, however, we are unable to find any fixed points with all positive components.

 Relaxing constraint (b), we can consider the matrix $\left(
\begin{array}{cc}
 0 & 1 \\
 1 & 0 \\
\end{array}
\right)$. This has fixed points at $(n,\frac 1n)^T$ for all $n\in \R_+$. There is a positive fixed point, but it is not unique. If we look at a matrix that is only a slight perturbation of this one, $\left(
\begin{array}{cc}
 0 & 1 +\epsilon \\
 1 & 0 \\
\end{array}
\right)$ for any small $\epsilon \in \R_+$, we find it has no positive fixed points at all.

This observation is what has motivated us, in the new packing heuristic's definition, to regard each set as a neighbour of itself. Without this, it is possible to formulate Set Packing instances for which the new packing heuristic has no consistent valuations. We have also considered a class of heuristics where we obtain valuations for the definition 
\[
\vecv = \matC ((bin(\matA^T \matA) - (1-\epsilon)\matI)\vecv)^\hinv
\]
for arbitrarily small positive epsilon. Matrices of this form satisfy the preconditions for our fixed point existence proof, and the iteration described in section \ref{sec:Calculating Fixed Points In Practice} does tend to find a fixed point. Using these fixed points for valuations in the greedy scheme appears to generate packings with high quality, but they are not obviously superior to the packings found by the new packing heuristic as presented in section \ref{sec:The New Greedy Set Packing Heuristic}. This looks to us like a potentially valuable direction for further investigation.

We have not found any explicit matrix $\matM$ that satisfies (a) and (b) but not (c) and is known to have more than one positive fixed point. The matrix $\matM$ used in our new packing heuristic are generally of this type. We believe it is possible that our new packing heuristic generates unique valuations, but do not have much confidence in either possible resolution. In the event that our packing heuristic produces unique valuations, we would regard it as more natural, but in either case the results of section \ref{sec:Set Packing Results} indicate that our heuristic produces high quality packings relative to alternative greedy packing heuristics.

\chapter{Experimental Results}
\label{chap:Experimental Results}

%%%%%%%%%%%%%%%%%%%%%%%%%%%%%%%%%%%%%%%%%%%%%%%
\section{Random Instances}
%%%%%%%%%%%%%%%%%%%%%%%%%%%%%%%%%%%%%%%%%%%%%%%
In order to test the heuristics we have described, we need to run the greedy algorithm on particular instances. To this end, we define a distribution of random instances. Let $\calD(m,n,\rho,C)$ where $m\in \N$ represents the number of sets, $n\in\N$ the number of elements, $\rho \in (0,1)$ the \emph{density} of the instance and $C$ the distribution of the set costs. We use $\calD(m,n,\rho,C)$ to represent the distribution of instances made by the following process:
\begin{enumerate}
\item
Fix $\calI=\set{1,\ldots,m}$
\item
For each $i\in\calI$, set $c_i$ to a sample drawn uniformly at random from $C$. 
\item
For each $i\in\calI$ set $S_i$, to a subset of $\set{1,\ldots,n}$ with each element selected independently with probability $\rho$.
\item
For each element $e\in\set{1,\ldots,n}$, if $e$ is in fewer than 2 input sets, add $e$ to $S_i$ for $i\in\set{1,\ldots,m}$ selected uniformly at random until $e$ is in 2 input sets.
\end{enumerate}

The cost distributions that we will consider are the following:
\begin{enumerate}
\item
$unweighted$. All costs are set to 1. In order to save space in tables of results, we will sometimes write $u$ instead of $unweighted$.
\item
$discrete(a,b)$ for $a,b\in \N$ with $0 < a < b$. All costs are selected with uniformly probability from the set $\set{a,\ldots,b}$. We will sometimes write $d(a,b)$ for this distribution.
\item
$continuous(a,b)$. All costs are selected with uniformly probability from the real interval $(a,b)$. We occasionally use $c(a,b)$ to denote this.
\end{enumerate}

We are interested in instances with these different distributions mainly because the standard heuristic is highly prone to ties in the unweighted case and only somewhat less so in the discrete cost setting. When the costs are random real numbers, the standard greedy algorithm is deterministic for all intents and purposes. When the standard valuation gives ties for some elements, we select the set to include at random. We do the same for the new heuristics, but they obtain ties far less frequently, even for unweighted instances. In order to provide a fair comparison between standard greedy and the new algorithm, in many cases we give both algorithms approximately equal time by doing multiple independent runs of the standard greedy algorithm and using the best cover/packing that it finds. The ability to break ties in different ways is the only advantage obtained by running the standard algorithms multiple times. This permits them to effectively sample from the space of solutions that they could potentially return. The new algorithms would not usually obtain any benefit from multiple runs, since they produce ties so rarely.

%%%%%%%%%%%%%%%%%%%%%%%%%%%%%%%%%%%%%%%%%%%%%%%
\section{Algorithms Used}
%%%%%%%%%%%%%%%%%%%%%%%%%%%%%%%%%%%%%%%%%%%%%%%

%For both SCP and SPP we can sometimes obtain exact solutions. This is done by encoding the problem as an IP and using IBM's CPLEX. For the smaller instances we deal with, CPLEX can often find optimal solutions in around 10 seconds, but for the larger instances it becomes untenable for us to solve them exactly. When we describe approximation ratios obtained by solving a problem exactly, we use $OPT$ for the optimal cost of a solution.

In all of the experiments below, we use only basic preprocessing. We have found that the more involved preprocessing techniques are not clearly of value, so we do not consider them here. Further work is needed to evaluate their usefulness.

For all of the valuation techniques discussed here, we consider it a tie if the valuations of input sets differ by less than $10^{-7}$. In the event of a tie, we select an input set at random from the sets with maximum valuation.

%%%%%%%%%%%%%%%%%%%%%%%%%%%%%%%%%%%%%%%%%%%%%%%
\subsection{Set Cover Algorithms}
%%%%%%%%%%%%%%%%%%%%%%%%%%%%%%%%%%%%%%%%%%%%%%%

For SCP, the algorithms we are comparing all use the general greedy scheme as described in \ref{sec:General Greedy Scheme} so they differ only in how they compute the valuations of the sets at each step. The following Cover valuations are considered:
\begin{enumerate}
\item
The standard heuristic ($STD$). $\vecv = \matC^{-1}\matA^T \vecone$. For some tests, we run the standard heuristic many times. We denote the best cover found with $k$ independent runs of the standard algorithm by $STD_k$.

\item
The new heuristic ($NEWC(\gamma)$), with parameter $\gamma$. 

$\vecv = \matC^{-1} \matA^T \vecd$ for $\vecd$ such that 
$
\vecd = \left(  
\matA \matC^{\gamma-1} \matA^T \vecd
\right)^\hinv
$

Multiple input sets obtaining the same valuation is very rare, relative to the standard heuristic, so we only ever use a single run of the new algorithm.

%\item
%An alternative formulation of the new heuristic ($ALT(\gamma)$), in which we find $\vecv$ for which $\vecv = \matC^\hinv\matA^T \vecone$ and $\vecd$ with components given by $\vecd_j =  1 / ({ \max_{i \text{ s.t. } j \in S_i} c_i^\gamma \vecv_i})$ are a consistent valuation. (\undone Need to describe this in the body of the text)
\end{enumerate}
We have omitted many other possible algorithms. The main reason for this is that other simple algorithms (e.g. Primal/Dual, LP Rounding) do not appear to be competitive with the standard greedy algorithm, as can be seen from Gomes et al.'s experimental work in \cite{gomes2006experimental}.

Except when otherwise noted, we always minimize returned covers by the Wool and Grossman technique described in section \ref{Wool and Grossman}. In practice, this is far more beneficial for the standard algorithm than for the new algorithm.

%%%%%%%%%%%%%%%%%%%%%%%%%%%%%%%%%%%%%%%%%%%%%%%
\subsection{Set Packing Algorithms}
%%%%%%%%%%%%%%%%%%%%%%%%%%%%%%%%%%%%%%%%%%%%%%%

For Set Packing, we consider the following valuation techniques:
\begin{enumerate}
\item
A variation of the standard Set Cover heuristic ($STDP$), where we pick the set with greatest weight per element. The valuations are given by $\vecv = \matC (\matA^T \vecone)^\hinv$. When run multiple times, we indicate this by $STDP_k$ for $k$ independent runs.

\item
The heuristic valuing sets by their weight divided by the square root of their size, which we call the ($ROOT$). $\vecv = \matC (\matA^T \vecone)^{\diamond -\frac 12}$. Multiple runs are denoted by $ROOT_k$.

\item
The standard MIS heuristic ($MIS$). $\vecv = \matC ((bin(\matA^T\matA) - \matI) \vecone)^\hinv$. Multiple runs are denoted by $STDP_k$.

\item
The new heuristic ($NEWP$). We choose $\vecv \in \R^m_+$ such that $\vecv = \matC (bin(\matA^T\matA) \vecv)^\hinv$.

\end{enumerate}

%%%%%%%%%%%%%%%%%%%%%%%%%%%%%%%%%%%%%%%%%%%%%%%
\section{Set Cover Results}
%%%%%%%%%%%%%%%%%%%%%%%%%%%%%%%%%%%%%%%%%%%%%%%

%%%%%%%%%%%%%%%%%%%%%%%%%%%%%%%%%%%%%%%%%%%%%%%
\subsection{Varying $\gamma$ for the New Heuristic}
\label{sec:Varying gamma for the New Heuristic}
%%%%%%%%%%%%%%%%%%%%%%%%%%%%%%%%%%%%%%%%%%%%%%%

In order to justify the choice of $\gamma=-3$ for the remaining tests, we have tested a variety of different values of $\gamma$ on a variety of different random problem distributions. For each distribution, we generate 100 instances and solve them with each heuristic. We have compared the new algorithm for $\gamma$ ranging between 0 and -4.  The results are presented in tables \ref{tab:gamma variations best} and \ref{tab:gamma variations quality}. Unweighted distributions are not tested, because all values of $\gamma$ yield the same valuations when all costs are identical.

In table \ref{tab:gamma variations best} we show the proportion of the 100 instances that the new heuristic with the stated choice of $\gamma$ performed best on. Table \ref{tab:gamma variations quality} shows the \emph{quality} of the obtained solutions. It is calculated as the average, over the 100 instances, of the cost obtained by each algorithm divided  by the cost of the best solution found by any of the algorithm runs. It is effectively a proxy for approximation ratio, which we cannot compute because the instances are too large to be solved exactly.

It can be readily seen that $\gamma=-3$ is best among the algorithms both in terms of finding the smallest cover found most frequently and in terms of quality for a majority of distributions. It is interesting, however, that $\gamma=-3$ does not dominate any of the alternatives. It seems reasonable that running the new heuristic with different values of $\gamma$ can be used to obtain different solutions, enabling us to utilize multiple runs of the new algorithm in the same way that its tendency to tie makes multiple runs of the standard heuristic valuable. 

It is interesting that for most of the distributions, there are few instances for which the best cover is found by more than one of these algorithms. With $\calD(5000, 500, 0.02, continuous(1,50))$, for instance, the best cover was found by exactly one of our settings for $\gamma$. The fairly wide spread between heuristics finding the best solution also suggests that multiple runs with different $\gamma$ can be valuable in practice. 

The setting $\gamma=-3$ appears to be the best overall, but it should be noted that there is no reason that $\gamma$ must be an integer. Arbitrary real values of $\gamma$ yield alternative versions of the new cover heuristic, and it may well be that some other value between -2 and -4 performs better than -3. Further experiments in this direction may be valuable.

\begin{table}
\centering
\caption{Comparing the effectiveness of different values for $\gamma$ for the new cover heuristic. Each cell indicated what proportion of 100 instances drawn from that row's distribution that column's heuristic performed best on.}
\label{tab:gamma variations best}
    \begin{tabular}{|l|c|c|c|c|c|c|}
    \hline                       Distribution &            $\gamma=0$ & $\gamma=-1$ & $\gamma=-2$ & $\gamma=-3$ & $\gamma=-4$  \\ \hline
       $\calD(500, 50, 0.05, discrete(1,50))$ &               22\% &               27\% &    \hicell    75\% &               66\% &               57\% \\ \hline
     $\calD(1000, 200, 0.02, discrete(1,50))$ &                6\% &                9\% &               43\% &   \hicell     50\% &               27\% \\ \hline
     $\calD(1000, 200, 0.05, discrete(1,50))$ &                9\% &                9\% &               37\% &   \hicell     57\% &               49\% \\ \hline
     $\calD(2000, 200, 0.02, discrete(1,50))$ &                2\% &               12\% &    \hicell    56\% &               47\% &               29\% \\ \hline
     $\calD(3000, 300, 0.02, discrete(1,50))$ &                4\% &                7\% &               42\% &   \hicell     59\% &               35\% \\ \hline
     $\calD(5000, 500, 0.02, discrete(1,50))$ &                3\% &                5\% &               35\% &   \hicell     52\% &               33\% \\ \hline
     $\calD(5000, 500, 0.05, discrete(1,50))$ &               10\% &               21\% &               29\% &               52\% &   \hicell     55\% \\ \hline
      $\calD(5000, 500, 0.1, discrete(1,50))$ &               29\% &               54\% &               57\% &               63\% &   \hicell     69\% \\ \hline
   $\calD(10000, 1000, 0.02, discrete(1,50))$ &                3\% &                6\% &               27\% &   \hicell     50\% &               49\% \\ \hline
\hline
     $\calD(500, 50, 0.05, continuous(1,50))$ &               10\% &               17\% &    \hicell    45\% &               42\% &               31\% \\ \hline
   $\calD(1000, 200, 0.02, continuous(1,50))$ &                1\% &                9\% &    \hicell    43\% &               31\% &               19\% \\ \hline
   $\calD(1000, 200, 0.05, continuous(1,50))$ &                8\% &               14\% &    \hicell    28\% &               26\% &               26\% \\ \hline
   $\calD(2000, 200, 0.02, continuous(1,50))$ &                2\% &                6\% &               36\% &   \hicell     40\% &               18\% \\ \hline
   $\calD(3000, 300, 0.02, continuous(1,50))$ &                1\% &                3\% &               36\% &   \hicell     40\% &               20\% \\ \hline
   $\calD(5000, 500, 0.02, continuous(1,50))$ &                5\% &                6\% &               29\% &   \hicell     39\% &               21\% \\ \hline
   $\calD(5000, 500, 0.05, continuous(1,50))$ &                6\% &                8\% &               16\% &   \hicell     39\% &               31\% \\ \hline
    $\calD(5000, 500, 0.1, continuous(1,50))$ &               10\% &               16\% &               15\% &   \hicell     36\% &               34\% \\ \hline
 $\calD(10000, 1000, 0.02, continuous(1,50))$ &                1\% &                6\% &               21\% &   \hicell     40\% &               32\% \\ \hline
\hline
         Average over all instances           &               7.33\%&             13.06\%&             37.22\%&    \hicell  46.06\%&     35.28\% \\ \hline

\end{tabular}
\end{table}

\begin{table}
\centering
\caption{Comparing the effectiveness of different values for $\gamma$ for the new cover heuristic. Each cell indicated the average performance of the new heuristic for the stated column's value of $\gamma$ over 100 instances drawn from that row's distribution.}
\label{tab:gamma variations quality}
    \begin{tabular}{|l|c|c|c|c|c|c|}
    \hline                       Distribution &  $\gamma=0$ Q& $\gamma=-1$ Q& $\gamma=-2$ Q& $\gamma=-3$ Q& $\gamma=-4$ Q \\ \hline
     $\calD(500, 50, 0.05, d(1,50))$ &             1.0383&             1.0282&  \hicell    1.0078&             1.0082&     1.0171 \\ \hline
   $\calD(1000, 200, 0.02, d(1,50))$ &             1.0337&             1.0253&             1.0081&    \hicell  1.0068&     1.0123 \\ \hline
   $\calD(1000, 200, 0.05, d(1,50))$ &             1.0447&             1.0367&             1.0178&    \hicell  1.0116&     1.0152 \\ \hline
   $\calD(2000, 200, 0.02, d(1,50))$ &             1.0384&             1.0252&             1.0079&    \hicell  1.0077&     1.0149 \\ \hline
   $\calD(3000, 300, 0.02, d(1,50))$ &             1.0374&             1.0287&             1.0107&    \hicell  1.0056&     1.0117 \\ \hline
   $\calD(5000, 500, 0.02, d(1,50))$ &             1.0407&             1.0298&             1.0132&    \hicell  1.0079&     1.0119 \\ \hline
   $\calD(5000, 500, 0.05, d(1,50))$ &             1.0489&             1.0337&             1.0243&             1.0144&     1.0133 \hicell \\ \hline
    $\calD(5000, 500, 0.1, d(1,50))$ &             1.0436&             1.0251&             1.0222&             1.0164&     1.0133 \hicell \\ \hline
 $\calD(10000, 1000, 0.02, d(1,50))$ &             1.0463&             1.0309&             1.0173&             1.0091&     1.0082 \hicell \\ \hline
\hline
     $\calD(500, 50, 0.05, c(1,50))$ &             1.0428&             1.0286&             1.0132&    \hicell  1.0098&     1.0142 \\ \hline
   $\calD(1000, 200, 0.02, c(1,50))$ &             1.0337&             1.0237&  \hicell    1.0057&             1.0080&     1.0149 \\ \hline
   $\calD(1000, 200, 0.05, c(1,50))$ &             1.0442&             1.0288&             1.0186&    \hicell  1.0158&     1.0182 \\ \hline
   $\calD(2000, 200, 0.02, c(1,50))$ &             1.0420&             1.0319&             1.0099&    \hicell  1.0092&     1.0154 \\ \hline
   $\calD(3000, 300, 0.02, c(1,50))$ &             1.0432&             1.0317&  \hicell    1.0088&             1.0094&     1.0153 \\ \hline
   $\calD(5000, 500, 0.02, c(1,50))$ &             1.0420&             1.0327&             1.0141&    \hicell  1.0077&     1.0136 \\ \hline
   $\calD(5000, 500, 0.05, c(1,50))$ &             1.0522&             1.0372&             1.0217&    \hicell  1.0148&     1.0167 \\ \hline
    $\calD(5000, 500, 0.1, c(1,50))$ &             1.0472&             1.0347&             1.0308&             1.0200&     1.0173 \hicell \\ \hline
 $\calD(10000, 1000, 0.02, c(1,50))$ &             1.0430&             1.0290&             1.0145&    \hicell  1.0083&     1.0086 \\ \hline
\hline
Average over all instances           &             1.0424&             1.0301&             1.0148&    \hicell  1.0106&     1.0140 \\ \hline
    \end{tabular}
\end{table}

%%%%%%%%%%%%%%%%%%%%%%%%%%%%%%%%%%%%%%%%%%%%%%%
\subsection{Comparison Between the Standard and New Heuristics}
%%%%%%%%%%%%%%%%%%%%%%%%%%%%%%%%%%%%%%%%%%%%%%%

In order to compare the standard and new heuristics as fairly as possible, we have to examine a few different situations. We run both algorithms on a variety of different distributions, generating 100 problems for every row of the tables below. For each problem, we run the standard heuristic 50 times, taking the best solution it obtains, and the new heuristic once. We track the cost of the best solution obtained before minimization and also the cost of the best solution after minimization. 

Table \ref{tab:cover preminimize main table} summarizes our results for the 2 algorithms, $STDC_{50}$ and $NEWC(-3)$, considering the quality of the solutions obtained for the random problems. Both algorithms are run using only the basic preprocessing steps described in section \ref{sec:Basic Preprocessing}. The columns labelled $STDC_{50}$ and $NEWC(-3)$ show the percentage of the 100 problems for which each algorithm obtained the best solution. It is possible for the sum of these values to exceed 100\% if for some of the instances, both algorithms return a set with the same cost. Under the columns labelled \emph{Quality}, we calculate the average ratio relative to the best solution found (by either algorithm \emph{after} minimization) which we use as a proxy for approximation ratio, since many of these problems cannot be solved exactly within a reasonable period of time. Let $s_i$ be the cost of the best of 50 runs of $STDC$ on instance number $i$ and $t_i$ be the cost of the set returned by $NEWC(-3)$. Then the ``{$STDC_{50}$ Quality}'' column contains the value $\frac{1}{100}\sum_{i=1}^{100} \frac{s_i}{\min(s_i,t_i)}$ and ``{$NEWC(-3)$ Quality}'' contains $\frac{1}{100}\sum_{i=1}^{100} \frac{t_i}{\min(s_i,t_i)}$. In every row the cell for the algorithm that performs best on the highest proportion of instances and the cell for the algorithm with best quality are highlighted.

\begin{table}
\centering
\caption{Comparison between the standard and new set cover heuristics for a range of random instance distributions. All results reported before the returned covers are minimized.
\label{tab:cover preminimize main table}}
\begin{tabular}{|l|c|c|c|c|}
\hline                                 Distribution &$STDC_{50}$  &$STDC_{50}$ Q&     $NEWC(-3)$ &     $NEWC(-3)$ Q \\ \hline
         $\calD(1000, 200, 0.02, unweighted)$ &   \hicell     74\%   &   \hicell   1.0085&               62\%   &             1.0138 \\ \hline
         $\calD(1000, 200, 0.05, unweighted)$ &   \hicell     99\%   &   \hicell   1.0015&               53\%   &             1.0244 \\ \hline
         $\calD(2000, 200, 0.02, unweighted)$ &   \hicell     91\%   &   \hicell   1.0033&               52\%   &             1.0170 \\ \hline
         $\calD(3000, 300, 0.02, unweighted)$ &   \hicell     97\%   &   \hicell   1.0013&               42\%   &             1.0186 \\ \hline
         $\calD(5000, 500, 0.02, unweighted)$ &   \hicell     95\%   &   \hicell   1.0010&               36\%   &             1.0157 \\ \hline
         $\calD(5000, 500, 0.05, unweighted)$ &   \hicell     99\%   &   \hicell   1.0007&               42\%   &             1.0239 \\ \hline
          $\calD(5000, 500, 0.1, unweighted)$ &   \hicell    100\%   &   \hicell   1.0000&               46\%   &             1.0318 \\ \hline
       $\calD(10000, 1000, 0.02, unweighted)$ &   \hicell     99\%   &   \hicell   1.0002&               28\%   &             1.0153 \\ \hline
\hline
     $\calD(1000, 200, 0.02, discrete(1,50))$ &                0\%   &             1.0749&   \hicell    100\%   &   \hicell   1.0025 \\ \hline
     $\calD(1000, 200, 0.05, discrete(1,50))$ &                2\%   &             1.0734&   \hicell     98\%   &   \hicell   1.0082 \\ \hline
     $\calD(2000, 200, 0.02, discrete(1,50))$ &                0\%   &             1.0711&   \hicell    100\%   &   \hicell   1.0047 \\ \hline
     $\calD(3000, 300, 0.02, discrete(1,50))$ &                0\%   &             1.0708&   \hicell    100\%   &   \hicell   1.0042 \\ \hline
     $\calD(5000, 500, 0.02, discrete(1,50))$ &                1\%   &             1.0646&   \hicell    100\%   &   \hicell   1.0044 \\ \hline
     $\calD(5000, 500, 0.05, discrete(1,50))$ &               39\%   &             1.0363&   \hicell     82\%   &   \hicell   1.0186 \\ \hline
      $\calD(5000, 500, 0.1, discrete(1,50))$ &   \hicell     83\%   &   \hicell   1.0147&               70\%   &             1.0196 \\ \hline
   $\calD(10000, 1000, 0.02, discrete(1,50))$ &                1\%   &             1.0481&   \hicell    100\%   &   \hicell   1.0066 \\ \hline
\hline
   $\calD(1000, 200, 0.02, continuous(1,50))$ &                0\%   &             1.1000&   \hicell    100\%   &   \hicell   1.0015 \\ \hline
   $\calD(1000, 200, 0.05, continuous(1,50))$ &                0\%   &             1.1036&   \hicell    100\%   &   \hicell   1.0026 \\ \hline
   $\calD(2000, 200, 0.02, continuous(1,50))$ &                0\%   &             1.0868&   \hicell    100\%   &   \hicell   1.0003 \\ \hline
   $\calD(3000, 300, 0.02, continuous(1,50))$ &                0\%   &             1.0911&   \hicell    100\%   &   \hicell   1.0012 \\ \hline
   $\calD(5000, 500, 0.02, continuous(1,50))$ &                0\%   &             1.0833&   \hicell    100\%   &   \hicell   1.0024 \\ \hline
   $\calD(5000, 500, 0.05, continuous(1,50))$ &                2\%   &             1.0652&   \hicell     98\%   &   \hicell   1.0048 \\ \hline
    $\calD(5000, 500, 0.1, continuous(1,50))$ &               15\%   &             1.0448&   \hicell     85\%   &   \hicell   1.0065 \\ \hline
 $\calD(10000, 1000, 0.02, continuous(1,50))$ &                1\%   &             1.0679&   \hicell     99\%   &   \hicell   1.0030 \\ \hline
\hline
  Average over all instances                  &               37.4\% &             1.0464&   \hicell     78.9\% &   \hicell   1.0105 \\ \hline

\end{tabular}
\end{table}

It can be seen that on the majority of weighted problems the new heuristic generally performs better regardless of instance size, though the standard heuristic performs better on some instances. The reason the standard algorithm works well on unweighted instances is that they have more situations where the standard valuation gives ties, allowing the 50 runs allocated to the standard algorithm to explore a variety of the possible solutions accessible to it. For the problems with continuous cost distributions, the standard heuristic is effectively deterministic, since the likelihood of ties occurring is very low. 

In table \ref{tab:cover postminimize main table} we show the results after minimizing all returned covers. The new heuristic fares only a little more poorly, though uniformly so. In table \ref{tab:cover quality difference main table} we show the difference in quality between the pre-minimized and the minimized solutions. Minimizing the covers returned by the new heuristic does not substantially reduce their sizes, improving them by around $0.1\%$ on average. 
It is effectively built into the new heuristic to avoid making selections that will later be made wholly redundant. This is not so for the standard heuristic. It is very common that the best minimized solution returned by the standard heuristic is significantly better than the best non-minimized solution it obtains. On average, the best minimized solution is $2.8\%$ better than the best non-minimized solution. 
For this reason, we believe that the new heuristic can be valuable for situations in which the minimization step is not possible, as in some formulations of online Set Cover or Hitting Set problems. One scheme that we suspect the new heuristic is particularly well suited for is model $\mathcal M2$ described in \cite{ausiello2009greedy}.

\begin{table}
\centering
\caption{Comparison between the standard and new set cover heuristics for a range of random instance distributions. All results reported after the returned covers are minimized.
\label{tab:cover postminimize main table}}
\begin{tabular}{|l|c|c|c|c|}
\hline                                 Distribution &$STDC_{50}$  &$STDC_{50}$  Q&     $NEWC(-3)$ &     $NEWC(-3)$ Q \\ \hline
         $\calD(1000, 200, 0.02, unweighted)$ &   \hicell     74\%   &   \hicell   1.0080&               62\%   &             1.0130 \\ \hline
         $\calD(1000, 200, 0.05, unweighted)$ &   \hicell     99\%   &   \hicell   1.0005&               52\%   &             1.0244 \\ \hline
         $\calD(2000, 200, 0.02, unweighted)$ &   \hicell     92\%   &   \hicell   1.0027&               51\%   &             1.0170 \\ \hline
         $\calD(3000, 300, 0.02, unweighted)$ &   \hicell     97\%   &   \hicell   1.0010&               42\%   &             1.0186 \\ \hline
         $\calD(5000, 500, 0.02, unweighted)$ &   \hicell     95\%   &   \hicell   1.0010&               36\%   &             1.0157 \\ \hline
         $\calD(5000, 500, 0.05, unweighted)$ &   \hicell     99\%   &   \hicell   1.0004&               41\%   &             1.0239 \\ \hline
          $\calD(5000, 500, 0.1, unweighted)$ &   \hicell    100\%   &   \hicell   1.0000&               46\%   &             1.0318 \\ \hline
       $\calD(10000, 1000, 0.02, unweighted)$ &   \hicell     99\%   &   \hicell   1.0002&               28\%   &             1.0153 \\ \hline
\hline
     $\calD(1000, 200, 0.02, discrete(1,50))$ &               27\%   &             1.0169&   \hicell     77\%   &   \hicell   1.0025 \\ \hline
     $\calD(1000, 200, 0.05, discrete(1,50))$ &               55\%   &             1.0117&   \hicell     71\%   &   \hicell   1.0078 \\ \hline
     $\calD(2000, 200, 0.02, discrete(1,50))$ &               40\%   &             1.0117&   \hicell     73\%   &   \hicell   1.0044 \\ \hline
     $\calD(3000, 300, 0.02, discrete(1,50))$ &               39\%   &             1.0109&   \hicell     77\%   &   \hicell   1.0037 \\ \hline
     $\calD(5000, 500, 0.02, discrete(1,50))$ &               38\%   &             1.0131&   \hicell     80\%   &   \hicell   1.0029 \\ \hline
     $\calD(5000, 500, 0.05, discrete(1,50))$ &   \hicell     63\%   &   \hicell   1.0099&               57\%   &             1.0140 \\ \hline
      $\calD(5000, 500, 0.1, discrete(1,50))$ &   \hicell     89\%   &   \hicell   1.0049&               58\%   &             1.0180 \\ \hline
   $\calD(10000, 1000, 0.02, discrete(1,50))$ &               20\%   &             1.0167&   \hicell     90\%   &   \hicell   1.0015 \\ \hline
\hline
   $\calD(1000, 200, 0.02, continuous(1,50))$ &                9\%   &             1.0357&   \hicell     91\%   &   \hicell   1.0012 \\ \hline
   $\calD(1000, 200, 0.05, continuous(1,50))$ &               10\%   &             1.0435&   \hicell     90\%   &   \hicell   1.0017 \\ \hline
   $\calD(2000, 200, 0.02, continuous(1,50))$ &                2\%   &             1.0381&   \hicell     98\%   &   \hicell   1.0002 \\ \hline
   $\calD(3000, 300, 0.02, continuous(1,50))$ &                7\%   &             1.0420&   \hicell     93\%   &   \hicell   1.0007 \\ \hline
   $\calD(5000, 500, 0.02, continuous(1,50))$ &                3\%   &             1.0443&   \hicell     97\%   &   \hicell   1.0005 \\ \hline
   $\calD(5000, 500, 0.05, continuous(1,50))$ &                7\%   &             1.0438&   \hicell     93\%   &   \hicell   1.0019 \\ \hline
    $\calD(5000, 500, 0.1, continuous(1,50))$ &               19\%   &             1.0362&   \hicell     81\%   &   \hicell   1.0035 \\ \hline
 $\calD(10000, 1000, 0.02, continuous(1,50))$ &                2\%   &             1.0446&   \hicell     98\%   &   \hicell   1.0004 \\ \hline
\hline
  Average over all instances                  &               49.4\% &             1.0182&   \hicell     70.1\% &   \hicell   1.0094 \\ \hline

\end{tabular}
\end{table}

Overall, we can see that the new heuristic is usually superior for instances where the set costs are drawn from a continuous distribution, but the results are mixed for distributions where the costs are uniform or discrete.

\begin{table}[!ht]
\centering
\caption{Quality difference between cover solutions before and after minimization.
\label{tab:cover quality difference main table}}
\begin{tabular}{|l|c|c|c|c|}
\hline                           Distribution &$STDC_{50}$ $\Delta$ Quality& $NEWC(-3)$ $\Delta$ Quality \\ \hline
         $\calD(1000, 200, 0.02, unweighted)$ &             0.0005&             0.0008 \\ \hline
         $\calD(1000, 200, 0.05, unweighted)$ &             0.0010&             0.0000 \\ \hline
         $\calD(2000, 200, 0.02, unweighted)$ &             0.0006&             0.0000 \\ \hline
         $\calD(3000, 300, 0.02, unweighted)$ &             0.0003&             0.0000 \\ \hline
         $\calD(5000, 500, 0.02, unweighted)$ &             0.0000&             0.0000 \\ \hline
         $\calD(5000, 500, 0.05, unweighted)$ &             0.0003&             0.0000 \\ \hline
          $\calD(5000, 500, 0.1, unweighted)$ &             0.0000&             0.0000 \\ \hline
       $\calD(10000, 1000, 0.02, unweighted)$ &             0.0000&             0.0000 \\ \hline
\hline
     $\calD(1000, 200, 0.02, discrete(1,50))$ &             0.0580&             0.0000 \\ \hline
     $\calD(1000, 200, 0.05, discrete(1,50))$ &             0.0617&             0.0004 \\ \hline
     $\calD(2000, 200, 0.02, discrete(1,50))$ &             0.0594&             0.0003 \\ \hline
     $\calD(3000, 300, 0.02, discrete(1,50))$ &             0.0599&             0.0005 \\ \hline
     $\calD(5000, 500, 0.02, discrete(1,50))$ &             0.0515&             0.0015 \\ \hline
     $\calD(5000, 500, 0.05, discrete(1,50))$ &             0.0264&             0.0046 \\ \hline
      $\calD(5000, 500, 0.1, discrete(1,50))$ &             0.0098&             0.0016 \\ \hline
   $\calD(10000, 1000, 0.02, discrete(1,50))$ &             0.0314&             0.0051 \\ \hline
\hline
   $\calD(1000, 200, 0.02, continuous(1,50))$ &             0.0643&             0.0003 \\ \hline
   $\calD(1000, 200, 0.05, continuous(1,50))$ &             0.0601&             0.0009 \\ \hline
   $\calD(2000, 200, 0.02, continuous(1,50))$ &             0.0487&             0.0001 \\ \hline
   $\calD(3000, 300, 0.02, continuous(1,50))$ &             0.0491&             0.0005 \\ \hline
   $\calD(5000, 500, 0.02, continuous(1,50))$ &             0.0390&             0.0019 \\ \hline
   $\calD(5000, 500, 0.05, continuous(1,50))$ &             0.0214&             0.0029 \\ \hline
    $\calD(5000, 500, 0.1, continuous(1,50))$ &             0.0086&             0.0030 \\ \hline
 $\calD(10000, 1000, 0.02, continuous(1,50))$ &             0.0233&             0.0026 \\ \hline
\hline
  Average over all instances                  &             0.0281&             0.0011 \\ \hline

\end{tabular}
\end{table}

\subsection{OR Library Instances}
\label{sec:OR Library Instances Cover}
%%%%%%%%%%%%%%%%%%%%%%%%%%%%%%%%%%%%%%%%%%%%%%%

The OR Library is a collection of optimization problems maintained by J.E. Beasley at \url{http://people.brunel.ac.uk/~mastjjb/jeb/info.html}. The Set Cover problems that we will be approximating are described at \url{http://people.brunel.ac.uk/~mastjjb/jeb/orlib/scpinfo.html}. They have perviously been used in experiments with SCP approximation in \cite{grossman1997computational}, and \cite{gomes2006experimental}. All of these problems have set costs chosen from $discrete(1,100)$.

The results are shown in table \ref{tbl:Cover OR Library instances}. The new heuristic has performance modestly better than 50 runs of the standard heuristic over these instances. We find the new heuristic to obtain the best solution in 60.3\% of the instances, and the standard heuristic only 54\%.

{\scriptsize
\begin{longtable}{|l|l|l|l|l|l|}
\caption{\normalsize A comparison of the approximate solutions obtained by the standard and new Set Cover heuristics for the OR Library instances.
\label{tbl:Cover OR Library instances}
}\\
\hline
  Instance Name &  $m$ &  $n$ & $\rho$   &$STDC_{50}$ & $NEWC(-3)$  \\ \hline
\endfirsthead				              
\caption{(continued)} \\		              
\hline					              
  Instance Name &  $m$ &  $n$ & $\rho$   &$STDC_{50}$ & $NEWC(-3)$  \\ \hline
\endhead				              
\endfoot				              
\endlastfoot
          scp41 & 1000 &  200 &   2.00\% &\hicell434 &        436 \\ \hline
          scp42 & 1000 &  200 &   1.99\% &       529 & \hicell513 \\ \hline
          scp43 & 1000 &  200 &   1.99\% &       537 & \hicell526 \\ \hline
          scp44 & 1000 &  200 &   2.00\% &\hicell504 &        512 \\ \hline
          scp45 & 1000 &  200 &   1.97\% &       518 & \hicell514 \\ \hline
          scp46 & 1000 &  200 &   2.04\% &       585 & \hicell565 \\ \hline
          scp47 & 1000 &  200 &   1.96\% &       447 & \hicell438 \\ \hline
          scp48 & 1000 &  200 &   2.01\% &       502 & \hicell493 \\ \hline
          scp49 & 1000 &  200 &   1.98\% &       663 & \hicell659 \\ \hline
         scp410 & 1000 &  200 &   1.95\% &       521 & \hicell516 \\ \hline
          scp51 & 2000 &  200 &   2.00\% &       269 & \hicell259 \\ \hline
          scp52 & 2000 &  200 &   2.00\% &       323 & \hicell318 \\ \hline
          scp53 & 2000 &  200 &   2.00\% &\hicell230 & \hicell230 \\ \hline
          scp54 & 2000 &  200 &   1.98\% &       247 & \hicell245 \\ \hline
          scp55 & 2000 &  200 &   1.96\% &\hicell212 & \hicell212 \\ \hline
          scp56 & 2000 &  200 &   2.00\% &       225 & \hicell218 \\ \hline
          scp57 & 2000 &  200 &   2.01\% &       301 & \hicell299 \\ \hline
          scp58 & 2000 &  200 &   1.98\% &       300 & \hicell294 \\ \hline
          scp59 & 2000 &  200 &   1.97\% &       290 & \hicell281 \\ \hline
         scp510 & 2000 &  200 &   2.00\% &       273 & \hicell272 \\ \hline
          scp61 & 1000 &  200 &   4.92\% &\hicell142 &        143 \\ \hline
          scp62 & 1000 &  200 &   5.00\% &       153 & \hicell150 \\ \hline
          scp63 & 1000 &  200 &   4.96\% &\hicell148 &        149 \\ \hline
          scp64 & 1000 &  200 &   4.93\% &       135 & \hicell134 \\ \hline
          scp65 & 1000 &  200 &   4.97\% &       178 & \hicell169 \\ \hline
          scpa1 & 3000 &  300 &   2.01\% &       259 & \hicell258 \\ \hline
          scpa2 & 3000 &  300 &   2.01\% &       264 & \hicell257 \\ \hline
          scpa3 & 3000 &  300 &   2.01\% &\hicell239 &        240 \\ \hline
          scpa4 & 3000 &  300 &   2.01\% &       240 & \hicell237 \\ \hline
          scpa5 & 3000 &  300 &   2.01\% &\hicell240 &        242 \\ \hline
          scpb1 & 3000 &  300 &   4.99\% & \hicell70 &         73 \\ \hline
          scpb2 & 3000 &  300 &   4.99\% & \hicell77 &         78 \\ \hline
          scpb3 & 3000 &  300 &   4.99\% & \hicell81 &         82 \\ \hline
          scpb4 & 3000 &  300 &   4.99\% &        83 &  \hicell82 \\ \hline
          scpb5 & 3000 &  300 &   4.99\% & \hicell72 &         73 \\ \hline
          scpc1 & 4000 &  400 &   2.00\% &       236 & \hicell234 \\ \hline
          scpc2 & 4000 &  400 &   2.00\% &\hicell224 & \hicell224 \\ \hline
          scpc3 & 4000 &  400 &   2.00\% &\hicell248 &        251 \\ \hline
          scpc4 & 4000 &  400 &   2.00\% &       231 & \hicell225 \\ \hline
          scpc5 & 4000 &  400 &   2.00\% &       220 & \hicell219 \\ \hline
          scpd1 & 4000 &  400 &   5.01\% & \hicell62 &         63 \\ \hline
          scpd2 & 4000 &  400 &   5.01\% & \hicell68 &  \hicell68 \\ \hline
          scpd3 & 4000 &  400 &   5.01\% & \hicell73 &         75 \\ \hline
          scpd4 & 4000 &  400 &   5.00\% & \hicell63 &  \hicell63 \\ \hline
          scpd5 & 4000 &  400 &   5.00\% & \hicell62 &         63 \\ \hline
        scpnre1 & 5000 &  500 &   9.98\% & \hicell29 &         30 \\ \hline
        scpnre2 & 5000 &  500 &   9.97\% & \hicell31 &         33 \\ \hline
        scpnre3 & 5000 &  500 &   9.97\% & \hicell28 &  \hicell28 \\ \hline
        scpnre4 & 5000 &  500 &   9.97\% & \hicell30 &         31 \\ \hline
        scpnre5 & 5000 &  500 &   9.98\% &        30 &  \hicell29 \\ \hline
        scpnrf1 & 5000 &  500 &  19.97\% & \hicell15 &  \hicell15 \\ \hline
        scpnrf2 & 5000 &  500 &  19.97\% & \hicell15 &         16 \\ \hline
        scpnrf3 & 5000 &  500 &  19.97\% & \hicell15 &         16 \\ \hline
        scpnrf4 & 5000 &  500 &  19.97\% & \hicell15 &  \hicell15 \\ \hline
        scpnrf5 & 5000 &  500 &  19.97\% & \hicell14 &  \hicell14 \\ \hline
        scpnrg1 &10000 & 1000 &   1.99\% &\hicell184 &        186 \\ \hline
        scpnrg2 &10000 & 1000 &   1.99\% &\hicell161 &        162 \\ \hline
        scpnrg3 &10000 & 1000 &   1.99\% &\hicell175 &        177 \\ \hline
        scpnrg4 &10000 & 1000 &   1.99\% &\hicell178 &        179 \\ \hline
        scpnrg5 &10000 & 1000 &   1.99\% &       179 & \hicell173 \\ \hline
        scpnrh1 &10000 & 1000 &   4.99\% & \hicell67 &         70 \\ \hline
        scpnrh2 &10000 & 1000 &   4.99\% &        68 &  \hicell66 \\ \hline
        scpnrh3 &10000 & 1000 &   4.99\% & \hicell63 &         64 \\ \hline
\hline
Best overall    &        & &  &       54.0\% & \hicell 60.3\% \\ \hline     
\end{longtable}
}

%%%%%%%%%%%%%%%%%%%%%%%%%%%%%%%%%%%%%%%%%%%%%%%
\section{Set Packing Results}
\label{sec:Set Packing Results}
%%%%%%%%%%%%%%%%%%%%%%%%%%%%%%%%%%%%%%%%%%%%%%%

%%%%%%%%%%%%%%%%%%%%%%%%%%%%%%%%%%%%%%%%%%%%%%%
\subsection{Comparison Between Packing Heuristics}
%%%%%%%%%%%%%%%%%%%%%%%%%%%%%%%%%%%%%%%%%%%%%%%

For this experiment, we run all 4 Set Packing heuristics on 100 instances from each of a variety of different random problem distributions. We use basic preprocessing for all algorithms and run all but the new heuristic 50 times, taking the best packing found in any of those runs as the result. Note that the running time of $MIS_{50}$ and $NEWP$ are comparable, but the running time of $STDP_{50}$ and $ROOT_{50}$ are significantly less than the running time of $NEWP$. The results of this experiment are shown in tables \ref{tab:Packing Main Results best percent} and \ref{tab:Packing Main Results quality}. Table \ref{tab:Packing Main Results best percent} shows what proportion of the instances for that row's distribution were solved with the largest weight packing among the 4 algorithms. Table \ref{tab:Packing Main Results quality} shows the quality of the packings produced. The quality is computed as the average, over all instances for that row, of the ratio between the best packing found and that algorithm's packing. This means that values nearer 1 are better. In both tables, we have highlighted the best achievement in each row.

\begin{table}[!ht]
\small
\centering
\caption{A comparison between the performance of 4 different heuristics for the Set Packing problem. The proportion of the 100 problems that each heuristic performed best on is reported.}
\label{tab:Packing Main Results best percent}
\begin{tabular}{|l|c|c|c|c|}
\hline                Distribution   &  $STDP_{50}$         &  $ROOT_{50}$         &  $MIS_{50}$          &  $NEWP$               \\ \hline
         $\calD(1000, 100, 0.02, unweighted)$ &    \hicell   100\%   &    \hicell   100\%   &   \hicell    100\%   &   \hicell    100\%    \\ \hline
         $\calD(1000, 200, 0.02, unweighted)$ &               44\%   &               44\%   &   \hicell     61\%   &               52\%    \\ \hline
         $\calD(1000, 200, 0.05, unweighted)$ &               69\%   &    \hicell    72\%   &               48\%   &               36\%    \\ \hline
         $\calD(2000, 200, 0.02, unweighted)$ &    \hicell    82\%   &               79\%   &               66\%   &               31\%    \\ \hline
         $\calD(3000, 300, 0.02, unweighted)$ &                8\%   &                5\%   &               62\%   &   \hicell     66\%    \\ \hline
         $\calD(5000, 500, 0.02, unweighted)$ &               17\%   &               27\%   &   \hicell     64\%   &               45\%    \\ \hline
         $\calD(5000, 500, 0.05, unweighted)$ &    \hicell    78\%   &               77\%   &               68\%   &               67\%    \\ \hline
          $\calD(5000, 500, 0.1, unweighted)$ &               62\%   &               62\%   &   \hicell     92\%   &               90\%    \\ \hline
       $\calD(10000, 1000, 0.02, unweighted)$ &    \hicell    62\%   &               55\%   &               43\%   &               27\%    \\ \hline
\hline
     $\calD(1000, 100, 0.02, discrete(1,50))$ &               17\%   &                7\%   &               16\%   &   \hicell     70\%    \\ \hline
     $\calD(1000, 200, 0.02, discrete(1,50))$ &                0\%   &                0\%   &                5\%   &   \hicell     95\%    \\ \hline
     $\calD(1000, 200, 0.05, discrete(1,50))$ &               11\%   &                9\%   &               15\%   &   \hicell     70\%    \\ \hline
     $\calD(2000, 200, 0.02, discrete(1,50))$ &                2\%   &                3\%   &                6\%   &   \hicell     91\%    \\ \hline
     $\calD(3000, 300, 0.02, discrete(1,50))$ &                2\%   &                5\%   &                5\%   &   \hicell     88\%    \\ \hline
     $\calD(5000, 500, 0.02, discrete(1,50))$ &                7\%   &                6\%   &               10\%   &   \hicell     77\%    \\ \hline
     $\calD(5000, 500, 0.05, discrete(1,50))$ &               17\%   &               25\%   &               32\%   &   \hicell     49\%    \\ \hline
      $\calD(5000, 500, 0.1, discrete(1,50))$ &               37\%   &               35\%   &               51\%   &   \hicell     55\%    \\ \hline
   $\calD(10000, 1000, 0.02, discrete(1,50))$ &               13\%   &               14\%   &               28\%   &   \hicell     47\%    \\ \hline
\hline
   $\calD(1000, 100, 0.02, continuous(1,50))$ &                8\%   &               10\%   &               19\%   &   \hicell     64\%    \\ \hline
   $\calD(1000, 200, 0.02, continuous(1,50))$ &                4\%   &                2\%   &                1\%   &   \hicell     93\%    \\ \hline
   $\calD(1000, 200, 0.05, continuous(1,50))$ &                4\%   &               16\%   &               19\%   &   \hicell     61\%    \\ \hline
   $\calD(2000, 200, 0.02, continuous(1,50))$ &                1\%   &                0\%   &                2\%   &   \hicell     97\%    \\ \hline
   $\calD(3000, 300, 0.02, continuous(1,50))$ &                2\%   &                0\%   &                1\%   &   \hicell     97\%    \\ \hline
   $\calD(5000, 500, 0.02, continuous(1,50))$ &                2\%   &                4\%   &                5\%   &   \hicell     89\%    \\ \hline
   $\calD(5000, 500, 0.05, continuous(1,50))$ &               15\%   &               20\%   &               19\%   &   \hicell     58\%    \\ \hline
    $\calD(5000, 500, 0.1, continuous(1,50))$ &               43\%   &               43\%   &               36\%   &   \hicell     45\%    \\ \hline
 $\calD(10000, 1000, 0.02, continuous(1,50))$ &                5\%   &               12\%   &               17\%   &   \hicell     66\%    \\ \hline
\hline
  Average over all instances         &              26.4\%  &               27.1\% &               33.0\% &   \hicell     67.6\%  \\ \hline

\end{tabular}
\end{table}

\begin{table}[!ht]
\small
\centering
\caption{A comparison between the performance of 4 different heuristics for the Set Packing problem. The average quality of each algorithm over 100 instances is reported.}
\label{tab:Packing Main Results quality}
\begin{tabular}{|l|c|c|c|c|}
\hline                Distribution   &  $STDP_{50}$ Q    &  $ROOT_{50}$ Q    &  $MIS_{50}$ Q     &  $NEWP$ Q          \\ \hline
         $\calD(1000, 100, 0.02, unweighted)$ &  \hicell    1.0000&   \hicell   1.0000&    \hicell  1.0000&  \hicell    1.0000 \\ \hline
         $\calD(1000, 200, 0.02, unweighted)$ &             1.0067&             1.0064&    \hicell  1.0041&             1.0061 \\ \hline
         $\calD(1000, 200, 0.05, unweighted)$ &             1.0122&   \hicell   1.0111&             1.0246&             1.0341 \\ \hline
         $\calD(2000, 200, 0.02, unweighted)$ &  \hicell    1.0013&             1.0015&             1.0029&             1.0074 \\ \hline
         $\calD(3000, 300, 0.02, unweighted)$ &             1.0176&             1.0184&    \hicell  1.0041&             1.0045 \\ \hline
         $\calD(5000, 500, 0.02, unweighted)$ &             1.0156&             1.0140&    \hicell  1.0062&             1.0112 \\ \hline
         $\calD(5000, 500, 0.05, unweighted)$ &  \hicell    1.0193&             1.0200&             1.0293&             1.0319 \\ \hline
          $\calD(5000, 500, 0.1, unweighted)$ &             1.1267&             1.1267&    \hicell  1.0267&             1.0333 \\ \hline
       $\calD(10000, 1000, 0.02, unweighted)$ &  \hicell    1.0094&             1.0118&             1.0189&             1.0295 \\ \hline
\hline
     $\calD(1000, 100, 0.02, discrete(1,50))$ &             1.0039&             1.0060&             1.0038&  \hicell    1.0005 \\ \hline
     $\calD(1000, 200, 0.02, discrete(1,50))$ &             1.0334&             1.0254&             1.0250&  \hicell    1.0003 \\ \hline
     $\calD(1000, 200, 0.05, discrete(1,50))$ &             1.0535&             1.0589&             1.0378&  \hicell    1.0088 \\ \hline
     $\calD(2000, 200, 0.02, discrete(1,50))$ &             1.0223&             1.0208&             1.0164&  \hicell    1.0003 \\ \hline
     $\calD(3000, 300, 0.02, discrete(1,50))$ &             1.0299&             1.0274&             1.0236&  \hicell    1.0007 \\ \hline
     $\calD(5000, 500, 0.02, discrete(1,50))$ &             1.0344&             1.0394&             1.0284&  \hicell    1.0025 \\ \hline
     $\calD(5000, 500, 0.05, discrete(1,50))$ &             1.0713&             1.0781&             1.0464&  \hicell    1.0235 \\ \hline
      $\calD(5000, 500, 0.1, discrete(1,50))$ &             1.0970&             1.1065&             1.0883&  \hicell    1.0548 \\ \hline
   $\calD(10000, 1000, 0.02, discrete(1,50))$ &             1.0435&             1.0458&             1.0269&  \hicell    1.0086 \\ \hline
\hline
   $\calD(1000, 100, 0.02, continuous(1,50))$ &             1.0053&             1.0065&             1.0039&  \hicell    1.0005 \\ \hline
   $\calD(1000, 200, 0.02, continuous(1,50))$ &             1.0290&             1.0247&             1.0253&  \hicell    1.0003 \\ \hline
   $\calD(1000, 200, 0.05, continuous(1,50))$ &             1.0633&             1.0553&             1.0415&  \hicell    1.0106 \\ \hline
   $\calD(2000, 200, 0.02, continuous(1,50))$ &             1.0265&             1.0246&             1.0198&  \hicell    1.0001 \\ \hline
   $\calD(3000, 300, 0.02, continuous(1,50))$ &             1.0356&             1.0352&             1.0294&  \hicell    1.0002 \\ \hline
   $\calD(5000, 500, 0.02, continuous(1,50))$ &             1.0479&             1.0435&             1.0335&  \hicell    1.0013 \\ \hline
   $\calD(5000, 500, 0.05, continuous(1,50))$ &             1.0660&             1.0587&             1.0471&  \hicell    1.0188 \\ \hline
    $\calD(5000, 500, 0.1, continuous(1,50))$ &  \hicell    1.0623&             1.0795&             1.1354&             1.1002 \\ \hline
 $\calD(10000, 1000, 0.02, continuous(1,50))$ &             1.0560&             1.0618&             1.0361&  \hicell    1.0062 \\ \hline
\hline
  Average over all instances         &             1.0367&             1.0373&             1.0291&  \hicell    1.0147 \\ \hline

\end{tabular}
\end{table}

Here, the new heuristic generally performs better than the alternatives. Much as with the new cover heuristic, it performs relatively poorly on the unweighted instances. This is again because the other algorithms see a larger range of possible solutions because of the multiple runs they are given. The results for the new packing heuristic are stronger than those for the cover heuristic. Here over two thirds of the time the new heuristic finds the best packing amongst all of the heuristics studied.

%%%%%%%%%%%%%%%%%%%%%%%%%%%%%%%%%%%%%%%%%%%%%%%
\subsection{OR Library Instances}
%%%%%%%%%%%%%%%%%%%%%%%%%%%%%%%%%%%%%%%%%%%%%%%

As in section \ref{sec:OR Library Instances Cover}, we have run the packing heuristics on the OR Library instances. The results are shown in table \ref{tbl:Packing OR Library instances}. 

The performance of the new heuristic over these instances is significantly better than the other heuristics, obtaining the best solution found over three quarters of the time. 

{\scriptsize
\begin{longtable}{|l|l|l|l|l|l|l|l|}
\caption{\normalsize A comparison of the approximate solutions obtained by the four set packing heuristics for the OR Library instances. 
\label{tbl:Packing OR Library instances}
}\\
\hline
  Instance Name &  $m$ &  $n$ & $\rho$ &$STD_{50}$ &$ROOT_{50}$  &$MIS_{50}$ &        $NEWP$  \\ \hline
\endfirsthead							                
\caption{(continued)} \\					                
\hline								                
  Instance Name &  $m$ &  $n$ & $\rho$ &$STD_{50}$  &$ROOT_{50}$  &$MIS_{50}$  &        $NEWP$  \\ \hline
\endhead							                
\endfoot							                
\endlastfoot
          scp41 & 1000 &   200 &  2.00\% &          5695 &          5639 &          5749 &   5887\hicell \\ \hline
         scp410 & 1000 &   200 &  1.95\% &          5940 &          6131 &          6066 &   6251\hicell \\ \hline
          scp42 & 1000 &   200 &  1.99\% &          5611 &          5747 &          5835 &   6044\hicell \\ \hline
          scp43 & 1000 &   200 &  1.99\% &          5996 &   6054\hicell &          5850 &          6035 \\ \hline
          scp44 & 1000 &   200 &  2.00\% &          5701 &          5808 &          5669 &   5905\hicell \\ \hline
          scp45 & 1000 &   200 &  1.97\% &          5746 &          5897 &          5949 &   6025\hicell \\ \hline
          scp46 & 1000 &   200 &  2.04\% &          6026 &          6136 &          6196 &   6234\hicell \\ \hline
          scp47 & 1000 &   200 &  1.96\% &          5949 &          6095 &          6040 &   6272\hicell \\ \hline
          scp48 & 1000 &   200 &  2.01\% &          6154 &          6152 &          6147 &   6378\hicell \\ \hline
          scp49 & 1000 &   200 &  1.98\% &          6393 &          6345 &          6360 &   6539\hicell \\ \hline
          scp51 & 2000 &   200 &  2.00\% &          8206 &          8197 &          8333 &   8459\hicell \\ \hline
         scp510 & 2000 &   200 &  2.00\% &          7716 &          7791 &          7724 &   7951\hicell \\ \hline
          scp52 & 2000 &   200 &  2.00\% &          7845 &          7964 &          7981 &   8155\hicell \\ \hline
          scp53 & 2000 &   200 &  2.00\% &          7750 &          7655 &          7874 &   7990\hicell \\ \hline
          scp54 & 2000 &   200 &  1.98\% &          8073 &          8091 &          8176 &   8212\hicell \\ \hline
          scp55 & 2000 &   200 &  1.96\% &          8030 &          8067 &          8038 &   8276\hicell \\ \hline
          scp56 & 2000 &   200 &  2.00\% &          7991 &          7956 &          8024 &   8144\hicell \\ \hline
          scp57 & 2000 &   200 &  2.01\% &          7841 &          7837 &          7748 &   7964\hicell \\ \hline
          scp58 & 2000 &   200 &  1.98\% &          7928 &          7901 &          7968 &   8139\hicell \\ \hline
          scp59 & 2000 &   200 &  1.97\% &          7918 &          8095 &          7993 &   8110\hicell \\ \hline
          scp61 & 1000 &   200 &  4.92\% &          1597 &          1571 &          1597 &   1625\hicell \\ \hline
          scp62 & 1000 &   200 &  5.00\% &          1693 &          1693 &   1815\hicell &          1774 \\ \hline
          scp63 & 1000 &   200 &  4.96\% &          1559 &          1597 &          1581 &   1797\hicell \\ \hline
          scp64 & 1000 &   200 &  4.93\% &   1910\hicell &          1753 &          1812 &   1910\hicell \\ \hline
          scp65 & 1000 &   200 &  4.97\% &   1627\hicell &          1457 &          1592 &          1618 \\ \hline
          scpa1 & 3000 &   300 &  2.01\% &          7410 &          7380 &          7580 &   7605\hicell \\ \hline
          scpa2 & 3000 &   300 &  2.01\% &          8000 &          7844 &          7880 &   8105\hicell \\ \hline
          scpa3 & 3000 &   300 &  2.01\% &          7483 &          7446 &          7422 &   7555\hicell \\ \hline
          scpa4 & 3000 &   300 &  2.01\% &          7699 &   7875\hicell &          7818 &          7871 \\ \hline
          scpa5 & 3000 &   300 &  2.01\% &          7510 &          7680 &          7568 &   7996\hicell \\ \hline
          scpb1 & 3000 &   300 &  4.99\% &          1453 &   1459\hicell &          1425 &          1440 \\ \hline
          scpb2 & 3000 &   300 &  4.99\% &          1419 &          1513 &   1571\hicell &          1545 \\ \hline
          scpb3 & 3000 &   300 &  4.99\% &          1360 &          1458 &          1447 &   1503\hicell \\ \hline
          scpb4 & 3000 &   300 &  4.99\% &          1314 &          1398 &          1413 &   1491\hicell \\ \hline
          scpb5 & 3000 &   300 &  4.99\% &          1437 &          1486 &          1519 &   1617\hicell \\ \hline
          scpc1 & 4000 &   400 &  2.00\% &          6684 &          6438 &          6669 &   6786\hicell \\ \hline
          scpc2 & 4000 &   400 &  2.00\% &          6076 &          6091 &   6242\hicell &          6224 \\ \hline
          scpc3 & 4000 &   400 &  2.00\% &          6246 &          6233 &          6260 &   6567\hicell \\ \hline
          scpc4 & 4000 &   400 &  2.00\% &          6652 &          6716 &          6579 &   6718\hicell \\ \hline
          scpc5 & 4000 &   400 &  2.00\% &          6794 &          6646 &          6565 &   6949\hicell \\ \hline
          scpd1 & 4000 &   400 &  5.01\% &   1141\hicell &          1036 &          1052 &          1131 \\ \hline
          scpd2 & 4000 &   400 &  5.01\% &          1084 &          1088 &          1144 &   1262\hicell \\ \hline
          scpd3 & 4000 &   400 &  5.01\% &          1183 &   1216\hicell &          1131 &          1200 \\ \hline
          scpd4 & 4000 &   400 &  5.00\% &          1195 &   1224\hicell &          1142 &   1224\hicell \\ \hline
          scpd5 & 4000 &   400 &  5.00\% &          1128 &          1145 &          1159 &   1201\hicell \\ \hline
        scpnre1 & 5000 &   500 &  9.98\% &           228 &    363\hicell &           296 &           296 \\ \hline
        scpnre2 & 5000 &   500 &  9.97\% &           224 &    359\hicell &           275 &           293 \\ \hline
        scpnre3 & 5000 &   500 &  9.97\% &    356\hicell &           241 &           274 &           274 \\ \hline
        scpnre4 & 5000 &   500 &  9.97\% &           221 &           239 &    295\hicell &    295\hicell \\ \hline
        scpnre5 & 5000 &   500 &  9.98\% &    354\hicell &           238 &           294 &           294 \\ \hline
        scpnrf1 & 5000 &   500 & 19.97\% &            95 &            95 &    100\hicell &    100\hicell \\ \hline
        scpnrf2 & 5000 &   500 & 19.97\% &            99 &            99 &    100\hicell &    100\hicell \\ \hline
        scpnrf3 & 5000 &   500 & 19.97\% &            99 &            99 &    100\hicell &    100\hicell \\ \hline
        scpnrf4 & 5000 &   500 & 19.97\% &            99 &            99 &    100\hicell &    100\hicell \\ \hline
        scpnrf5 & 5000 &   500 & 19.97\% &            99 &            99 &    100\hicell &    100\hicell \\ \hline
        scpnrg1 &10000 &  1000 &  1.99\% &          2922 &          2821 &   3097\hicell &          3035 \\ \hline
        scpnrg2 &10000 &  1000 &  1.99\% &          2993 &   3006\hicell &          2993 &          2953 \\ \hline
        scpnrg3 &10000 &  1000 &  1.99\% &          3100 &          3069 &          2872 &   3122\hicell \\ \hline
        scpnrg4 &10000 &  1000 &  1.99\% &          2946 &          2812 &          2987 &   3159\hicell \\ \hline
        scpnrg5 &10000 &  1000 &  1.99\% &          2942 &          2787 &          2980 &   3125\hicell \\ \hline
        scpnrh1 &10000 &  1000 &  4.99\% &    532\hicell &    532\hicell &    532\hicell &    532\hicell \\ \hline
        scpnrh2 &10000 &  1000 &  4.99\% &    531\hicell &    531\hicell &    531\hicell &    531\hicell \\ \hline
        scpnrh3 &10000 &  1000 &  4.99\% &    531\hicell &    531\hicell &    531\hicell &    531\hicell \\ \hline
\hline
  Average over all instances   &&&&              12.7\%  &         17.5\%&        20.6\% &    \hicell  76.2\%  \\ \hline

\end{longtable}
}

%%%%%%%%%%%%%%%%%%%%%%%%%%%%%%%%%%%%%%%%%%%%%%%%
%\subsection{Effectiveness of Preprocessing}
%%%%%%%%%%%%%%%%%%%%%%%%%%%%%%%%%%%%%%%%%%%%%%%%
%
%\workingnote{This test should be done again with all 4 algorithms.}
%
%We generated 200 instances from $\calD(500,100, 0.03, discrete(1,3))$ and use the same heuristic with different levels of preprocessing done. In one case, we do no preprocessing, in the second, only the basic preprocessing steps, and in the third, both basic preprocessing and subsumption preprocessing. We note what proportion of the 100 problems each algorithm performed the best of the three on and we note the mean approximation ratio attained by each algorithm.
%
%    \begin{tabular}{|l|l|l|l|l|}
%    \hline
%    ~                     & $NEW$ \% Best & $avg(OPT/NEW)$ & $ROOT$ \% Best & $avg(OPT/ROOT)$ \\ %\hline
%    No Preprocessing      & 63.5\%        & 1.0084         & 29\%           & 1.0201          \\ %\hline
%    Basic  Preprocessing  & 63\%          & 1.0084         & 54\%           & 1.0158          \\ %\hline
%    Subsumption + Basic   & 73.5\%        & 1.0067         & 79.5\%         & 1.0130          \\ %\hline
%    \end{tabular}

%%%%%%%%%%%%%%%%%%%%%%%%%%%%%%%%%%%%%%%%%%%%%%%
\chapter{Discussion}
%%%%%%%%%%%%%%%%%%%%%%%%%%%%%%%%%%%%%%%%%%%%%%%

We have demonstrated a novel method for devising greedy approximation heuristics. For two problems, Set Cover and Set Packing, we have constructed new heuristics and shown that their performance is better than alternative greedy algorithms. For the Set Cover heuristic, we have demonstrated that the valuation we define exists and is unique, while we are unsure whether the Set Packing heuristic guarantees a unique valuation. This leaves a variety of open questions. Is the valuation determined by the Packing heuristic unique? Is there some simple way to characterize matrices $\matM$ for which there is a unique solution $\vecv$ to $\matM\vecv = \vecv^\hinv$? Do the iterations we have described always converge? Is there a more efficient way to find these fixed points? Do either of the new heuristics guarantee some approximation ratio? Most importantly, can the overall technique of defining valuations recursively be used to construct high quality approximation algorithms for other problems? 

In our experimental results, it is striking that none of the algorithms considered appears to dominate the others. Although the new heuristics are generally preferable to the alternatives on the sorts of random instances we have used, it is still beneficial to use a variety of heuristics when looking for good covers and packings. We can straightforwardly recommend that any real-world software in which any of the standard cover or packing heuristics are used exclusively to obtain approximate solutions could substantially benefit from also considering the solutions given by our new heuristics. Although their runtimes are longer, they are not substantially so, and is should be possible to engineer them to run in only 5 or 10 times what's required for the standard heuristics. 

It is both a blessing and a curse that the new heuristics are less prone to ties than the standard heuristics. It is convenient that a single run deterministically generates a particular solution, but at the same time, this means that multiple runs will not allow us to sample from the space of possible solutions in the same way as the standard algorithms do. In order to generate a wider variety of solutions using the new heuristics there are a few different approaches that might be made. For the cover heuristic, varying the parameter $\gamma$ permits us to obtain different covers. Additionally, for any valuation-producing heuristic, we could randomize the general greedy scheme to select sets with probabilities determined by the valuations produced. We leave it to others to determine how valuable are the solutions built with this approach to obtaining variety.

%% This adds a line for the Bibliography in the Table of Contents.
\addcontentsline{toc}{chapter}{Bibliography}
%% *** Set the bibliography style. ***
%% (change according to your preference/requirements)
\bibliographystyle{plain}
%% *** Set the bibliography file. ***
\bibliography{ut-thesis}

%% *** NOTE ***
%% If you don't use bibliography files, comment out the previous line
%% and use \begin{thebibliography}...\end{thebibliography}.  (In that
%% case, you should probably put the bibliography in a separate file and
%% `\include' or `\input' it here).

\chapter{Acknowledgements}
%% *** Put your Acknowledgements here. ***

I'm grateful for the support and patience of my supervisors, Allan Borodin and Ken Jackson. I would also like to thank Danny Ferriera for many excellent conversations relating to this work, and Stephanie Southmayd for her kind help in editing this document. A great deal of thanks is due to Ken Jackson, who was the first to prove both the existence and uniqueness results that form the body of chapter \ref{chap:Mathematical Results}. I would also like to thank the OGS program and the University of Toronto for financial support.

\end{document}